\documentclass[11pt]{article}
\pdfoutput=1

\usepackage{stolenstyle}

\usepackage{amsmath,epsf,amssymb,latexsym,amsthm,setspace,bbm,array,pifont,enumerate}

\DeclareFontFamily{U}{rsf}{}
\DeclareFontShape{U}{rsf}{m}{n}{
  <5> <6> rsfs5 <7> <8> <9> rsfs7 <10-> rsfs10}{}
\DeclareMathAlphabet\Scr{U}{rsf}{m}{n}

\def\CO#1#2{{[#1,#2]}}

\def\GUL{\GU(1)_{\text{L}}}
\def\GUR{\GU(1)_{\text{R}}}

\def\C{{\mathbb C}}

\def\Q{{\mathbb Q}}
\def\P{{\mathbb P}}
\def\R{{\mathbb R}}
\def\Z{{\mathbb Z}}

\def\Hom{\operatorname{Hom}}

\def\Tr{\operatorname{Tr}}

\def\ch{\operatorname{ch}}

\def\im{\operatorname{im}}

\def\GL{\operatorname{GL}}

\def\GO{\operatorname{O{}}}
\def\SU{\operatorname{SU}}
\def\GU{\operatorname{U{}}}
\def\Sp{\operatorname{Sp}}

\def\la{\langle}
\def\ra{\rangle}

\def\ff#1#2{{\textstyle\frac{#1}{#2}}}

\def\cA{{\cal A}}

\def\cC{{\cal C}}

\def\cE{{\cal E}}

\def\cG{{\cal G}}
\def\cH{{\cal H}}

\def\cL{{\cal L}}
\def\cM{{\cal M}}
\def\cN{{\cal N}}
\def\cO{{\cal O}}

\def\cR{{\cal R}}
\def\cS{{\cal S}}

\newcommand\mub{\overline{\mu}}

\newcommand\bb{\overline{b}}
\newcommand\cb{\overline{c}}

\newcommand\gb{\overline{g}}

\newcommand\tb{\overline{t}}

\newcommand\pt{\widetilde{p}}

\newcommand\Gb{\overline{G}}

\newcommand\Jb{\overline{J}}

\newcommand\Lb{\overline{L}}

\newcommand\Xb{\overline{X}}
\newcommand\Yb{\overline{Y}}

\theoremstyle{definition}

\usepackage{tikz}

\usetikzlibrary[shapes.geometric]
\usetikzlibrary{positioning} 
\usetikzlibrary{calc,intersections,through,backgrounds}
\usetikzlibrary{cd}

\tikzset{>=stealth}
\tikzset{every picture/.style={very thick}}

\def\Ttor{{{\mathbb{T}}}}

\def\ba{{\boldsymbol{a}}}
\def\bb{{\boldsymbol{b}}}

\def\bq{{\textbf{q}}}

\def\ii{{\mathrm{i}}}

\def\sleft{\text{\tiny{L}}}
\def\sright{\text{\tiny{R}}}

\def\Ttor{{\boldsymbol{\mathbb{T}}}}

\def\preprint{ ~~ }

\title{The stringy geometry of integral cohomology in mirror symmetry}
\author[a] {Peng Cheng,}
\author[b] {Ilarion V.~Melnikov,}
\author[c] {Ruben Minasian}
\affiliation[a] {Arnold Sommerfeld Center, LMU M\"unchen, Theresienstra{\ss}e 37, 80333 M\"unchen, Germany}
\affiliation[b] {Department of Physics and Astronomy,
James Madison University,
Harrisonburg, VA 22807, USA}
\affiliation[c] {Institut de Physique Th{\'e}orique,  
Universit{\'e} Paris-Saclay, CNRS, CEA, F-9119, Gif-sur-Yvette, France
}

\emailAdd{Peng.Cheng@physik.uni-muenchen.de}
\emailAdd{melnikix@jmu.edu}
\emailAdd{ruben.minasian@ipht.fr}

\abstract{We examine the physical significance of torsion co-cycles in the cohomology of a projective Calabi-Yau three-fold for the (2,2) superconformal field theory (SCFT) associated to the non-linear sigma model with such a manifold as a target space.  There are two independent torsion subgroups in the cohomology. While one is associated to an orbifold construction of the SCFT, the other encodes the possibility of turning on a topologically non-trivial flat gerbe for the NS-NS B-field.  Inclusion of these data enriches  mirror symmetry by providing a refinement of the familiar structures and points to a generalization of the duality symmetry, where the topology of the flat gerbe enters on the same footing as the topology of the underlying manifold.
}

\begin{document}

\maketitle

\section{Introduction} \label{s:Introduction}

Consider the compactification of type IIA string theory on a smooth projective Calabi-Yau (CY) $3$-fold $X$.  It is a textbook fact that such a compactification preserves $8$ supercharges in $4$ dimensions, and its massless spectrum is characterized by the Hodge numbers $h^{1,2}(X)$ and $h^{1,1}(X)$.  In the same hypothetical textbook we also find the statement that there may be another CY $3$-fold $X^\circ$, with $h^{1,2}(X^\circ) = h^{1,1}(X)$ and $h^{1,1}(X^\circ) = h^{1,2}(X)$, such that compactification of IIB string theory on $X^\circ$ leads to an isomorphic theory in four dimensions.

At the heart of this remarkable duality is closed string mirror symmetry: an isomorphism
of two superconformal field theories: the (2,2) superconformal field theory (SCFT) $\cC[X]$ based on a non-linear sigma model (NLSM) with target space $X$ and another (2,2) SCFT $\cC[X^\circ]$ based on a NLSM with target space $X^\circ$.  The exchange of the Hodge numbers is the most basic and familiar aspect of mirror symmetry.  It has an immediate and satisfying reflection in the SCFT:  $h^{1,2}(X)$ encodes the number of exactly marginal operators with charges $q_{\sleft} = q_{\sright} = 1$ with respect to the $\GUL\times\GUR$ R-symmetry of the SCFT (these are the ``cc deformations''), while $h^{1,1}(X)$ encodes the exactly marginal operators with charges $-q_{\sleft} =q_{\sright} = 1$ (these are the ``ac deformations'').   Integrating these deformations we obtain the conformal manifold $\cM(X) = \cM_{\text{cc}}(X)\times\cM_{\text{ac}}(X)$ of the SCFT, where each factor is a special K\"ahler manifold with respect to the Zamolodchikov metric.

The present work is devoted to topological invariants encoded in the integral cohomology of $X$ beyond the Hodge numbers: in particular the torsion subgroups.  Poincar{\'e} duality and the universal coefficient theorem reduce these to two independent groups.  We follow~\cite{Batyrev:2005jc} and denote these by  
\begin{align}
A (X) &= \{H^2(X,\Z)\}_{\text{tor}}\quad \text{and} \quad B(X) = \{H^3(X,\Z)\}_{\text{tor}}~.
\end{align}  
Each of these has a clear role in the quantum field theory associated to the NSLM with target space $X$:  $A(X)$ encodes an abelian global symmetry present for all points in $\cM(X)$, while $B(X)$ encodes the possibility of turning on a topologically non-trivial gerbe for the $B$-field.  Our goal is to describe the role these groups play in closed string mirror duality, and our main conclusion is that their inclusion requires us to broaden the framework for the mirror correspondence: rather than being merely decorations on top of the textbook duality, their presence can lead to new dual pairs.

Before describing the physics of the closed string sector associated to $A(X)$ and $B(X)$, we point out what we might expect about their mirror duals.  The intuition from Calabi-Yau manifolds in higher dimension~\cite{Greene:1993vm}, as well as K3 mirror symmetry~\cite{MR1416354}, is that we should think of the mirror action not just on the Hodge numbers but rather on the total cohomology.  At the level of integral cohomology of CY $3$-folds this suggests the isomorphism of abelian groups $H^{\text{even}}(X,\Z) \simeq H^{\text{odd}}(X^\circ,\Z)$, which implies
\begin{align}
\label{eq:torsiongroupsmirror}
A(X) \oplus B(X)^\ast \simeq B(X^\circ) \oplus A(X^\circ)^\ast~,
\end{align}
where for any finite abelian group $G$ we denote its Pontryagin dual---defined below---by $G^\ast$.  We will mostly work with examples where $A(X) \simeq B(X^\circ)$ and $B(X) \simeq A(X^\circ)$ separately, and in fact one of the pairs is trivial, but such a refined isomorphism does not hold in general~\cite{Aspinwall:1995mh,Braun:2017oak}.  We will show that when every element of $A(X)$ has odd order~(\ref{eq:torsiongroupsmirror}) is a consequence of open mirror symmetry for $X$ and $X^\circ$.  This comes about through the relationship between $A(X)$ and $B(X)$ and the torsion subgroups  $\{K^0(X)\}_{\text{tor}}$ and $\{K^1(X)\}_{\text{tor}}$ of the K-theory groups, which classify the charges of torsional D-branes on $X$~\cite{Minasian:1997mm,Witten:1998cd,Brunner:2001eg,Brunner:2001sk}.

Returning to the closed string sector, the significance of $A(X)$ is fairly clear from its interpretation as the Pontryagin dual of the abelianization of the fundamental group $\pi_1(X)$.  Since we can construct $X$ as a quotient $\Xb /\Gamma$, where $\Xb$ is an isometric universal cover of $X$ that admits the free action of a group $\Gamma \simeq \pi_1(X)$, $\cC[X]$ is an orbifold conformal field theory:  $\cC[X] = \cC[\Xb]/\Gamma$.  If $X$ has a known mirror $X^\circ$ we can find a theory $\overline{\cC}^\circ$ with a global symmetry $\Gamma^\circ$ isomorphic to $\Gamma$ such that $\overline{\cC}^\circ/\Gamma^\circ = \cC[X^\circ]$.  When $\Gamma$ is finite and  abelian, this follows from standard properties of the quantum symmetry associated to an abelian orbifold, but for more general finite groups it requires a categorical symmetry construction---reviewed and axiomatized in~\cite{Brunner:2013xna,Bhardwaj:2017xup}.\footnote{If $\Gamma$ is of infinite order, then $X$ is a quotient of $T^6$ or $\text{K3}\times T^2$ by a freely--acting group, and the possibilities are classified in \cite{Hashimoto:2014oma}. These theories should be considered in its own right for many purposes, including mirror symmetry, which is now inherited from simpler operations in the cover~\cite{Hashimoto:2015zqm}.}   In either case, however, there is no guarantee that $\overline{\cC}^\circ$ can be presented as $\cC[\Yb]$ for some CY $\Yb$; as we show, even when it can be so presented, the action of the group $\Gamma^\circ$ may not have a geometric interpretation.

The relationship between $B(X)$ and the B-field flat gerbe is through the exact sequence
\begin{equation}
\label{eq:flatgerbecharacterization}
\begin{tikzcd}
0 \ar[r] & H^2(X,\R)/H^2(X,\Z) \ar[r] & H^2(X,\GU(1)) \ar[r] & B(X) \ar[r] & 0~.
\end{tikzcd}
\end{equation}
The middle term has the intuitive interpretation as a choice of phases assigned by the NLSM path integral to topological sectors labeled by the homology classes $[f(\Sigma)] \in H_2(X,\Z)$, where $f: \Sigma \to X$ is the map from the worldsheet $\Sigma$ to the target space $X$. Thus $B(X)$ describes the equivalence classes of flat gerbes with respect to shifts of $B$---the connection on the gerbe---by closed $2$-forms.  

Such flat but topologically non-trivial gerbes deserve attention for a number of reasons.  First, they are examples of flux vacua where stringy effects are under control because a flat B-field gerbe is naturally incorporated in the NLSM, and there is no issue in including it in a Ramond--Neveu-Schwarz formulation of the string worldsheet.  This should be contrasted with Ramond-Ramond (RR) flux backgrounds, for which it is notoriously difficult to incorporate stringy corrections.   RR fluxes, even when they they are flat, play an important and yet mysterious role in string duality~\cite{Ferrara:1995yx,Aspinwall:1998he}, as well as in fixing certain geometric parameters, for instance in the context of frozen singularities in F/M theory~\cite{deBoer:2001wca,Tachikawa:2015wka,Bhardwaj:2018jgp}.  In some situations a choice of such flat RR flux can be understood as the S-dual of a flat B-field gerbe~\cite{Cheng:2022nso}, and the freezing of the geometric parameters shown explicitly in the worldsheet theory.    We can therefore hope that lessons from flat gerbes, combined with string duality, can shed light on flat RR backgrounds.  In addition, as shown in the recent work~\cite{Katz:2022lyl,Katz:2023zan}, certain limiting points in $\cM_{\text{ac}}(X)$ can be described as a singular space $X'$ equipped with a flat B-field gerbe, and  the gerbe ensures that the SCFT remains smooth, despite the conifold singularities of $X'$.  Thus, flat gerbes arise naturally in the context of (2,2) SCFTs with a geometric interpretation.

The flat gerbe also plays a dramatic role in mirror symmetry.  Let us denote by $\cC[X,\beta]$, with $\beta \in B(X)$, the theory obtained by turning on a flat gerbe.  Given that $\cC[X^\circ]$ is isomorphic to $\cC[X]$,
is there an operation that can be performed on $\cC[X^\circ]$ to obtain the mirror of $\cC[X,\beta]$?  If such an operation exists, it may not be possible to describe it in terms of geometric or non-linear sigma model structures. For example, a mirror description of $\cC[X,\beta]$ may be of the form $\cC[Y^\circ,\beta^\circ]$, where $Y^\circ$ is topologically distinct from the original mirror $X^\circ$.  

The previous discussion involves two concepts in the original and mirror descriptions: orbifold theories related to a non-trivial $A(X)$, and flat gerbes related to a non-trivial $B(X)$.  There are intriguing and partially--understood relations between these notions.  For example, turning on certain flat gerbes in a theory with an orbifold presentation can be associated to a choice of discrete torsion in the orbifold construction.  This observation and its relation to mirror symmetry go back to~\cite{Vafa:1994rv}, and efforts to explore and extend it to general mirror pairs include~\cite{Aspinwall:1994uj,Kreuzer:1994qp,Aspinwall:1995rb,Kreuzer:1995yi,Sharpe:2000ki,Sharpe:2003cs,Batyrev:2005jc}.  We will use the relationship between the flat gerbe and a choice of discrete torsion to construct explicit mirror pairs $\cC[X,\beta]$ and $\cC[Y^\circ,\beta^\circ]$ in the context of the classic mirror construction of~\cite{Greene:1990ud}.  

The rest of this article is organized as follows.  In section~\ref{s:topogeometry} we review the essential topology and geometry underlying our subsequent discussion, and we show that
(\ref{eq:torsiongroupsmirror}) follows from the isomorphism of K-groups of the mirror manifolds.  Next, in section~\ref{s:orbifoldcovers} we tackle the discussion of $A(X)$ and its role in mirror symmetry.  Section~\ref{s:gerbesdt} deals with gerbes and mirror constructions with discrete torsion, and we conclude with a discussion and further directions in section~\ref{s:discussion}.

\section*{Acknowledgements} PC is partially supported by the DFG Excellence Strategy EXC-2094 390783311.  IVM's work is supported in part by the Humboldt Research Award and the Jean d'Alembert Program at the University of Paris--Saclay, as well as the Educational Leave program at James Madison University.  RM is partially supported by ERC grants 772408-Stringlandscape and 787320-QBH Structure, as well as the Humboldt Foundation.    We thank I.~Brunner, R.~Field, D.~Isra\"el,  M.R.~Plesser, T. Schimannek, and E.~Sharpe for useful discussions, and we are grateful to T.~H\"ubsch for pointing out a typo in a previous version.  This work initiated during a stay at the Albert Einstein Institute (Max Planck Institute for Gravitational Physics), and we are grateful to the AEI for its generous hospitality.  IVM also thanks the LPTHE at Sorbonne University for hospitality while this work was being completed.

\section{A review of topology and geometry}  \label{s:topogeometry}
Let $X$ be a Calabi-Yau manifold of complex dimension $d$.\footnote{There are a few slightly different definitions of Calabi-Yau manifolds, and we choose one that is general enough to cover the examples of interest to us, while excluding special cases like $T^6$.  Our presentation of the geometry mostly follows~\cite{Joyce:2000cm} and~\cite{Besse:1987pua}, to which we refer the reader for further discussion and references.}  That is, $X$ is a compact K\"ahler manifold with $H^i(X,\cO_X) = 0$ for $ 0<i<d$, while $H^0(X,\cO_X)$ and $H^d(X,\cO_X)$ are each isomorphic to $\C$.  It follows from Yau's theorem that $X$ admits a Ricci-flat metric $g$ with holonomy group $\text{Hol}(g) \subseteq \SU(d)$.   

We will now review some standard results from algebraic topology and geometry regarding these spaces that are often omitted in discussions of mirror symmetry.  Although no doubt known to the experts, it is perhaps useful for many readers (and certainly for the authors) to state these facts before we launch into an exploration of their significance in the NLSM, the SCFT, and in mirror duality.  

\subsection{Holonomy and the fundamental group} \label{ss:holofun}
The holonomy group $\text{Hol}(g)$ has a normal subgroup $\text{Hol}^0(g) \subset \text{Hol}(g)$ generated by the null-homotopic loops, and there is a surjective homomorphism $\pi_1(X) \to  \text{Hol}(g)/\text{Hol}^0(g)$.  The topology of $X$ is constrained by the Cheeger-Gromoll theorem---which applies to complete manifolds with non-negative Ricci curvature (a pedagogic presentation is given in~\cite{Besse:1987pua})---and its specialization to the case of K\"ahler Ricci-flat metrics~\cite{Beauville:1983skc}.  The  results relevant for us are as follows:
\begin{enumerate}
\item the fundamental group $\pi_1(X)$ has a finite normal subgroup $F$ such that $\pi_1(X) /F$ is a \textit{crystallographic group}, i.e. a discrete co-compact subgroup of $\R^k \rtimes \GO(k)$ for some $k\le \dim X$;\footnote{This statement is slightly different but equivalent to the one given in~\cite{Besse:1987pua}.  We find it more illuminating, as it directly shows the role of the isometry group of $\R^k$ in the universal cover of $X$.  Further details can be found in~\cite{Wilking:1999fg}.} 
\item when $(X,g)$ is K\"ahler and Ricci flat, its universal cover $(\Xb,\gb)$ is isomorphic, as a K\"ahler manifold, to a product $\C^k \times \prod_i Y_i \times \prod_j Z_j$, where each $(Y_i,\gb_i)$ is a compact K\"ahler simply-connected manifold with $\text{Hol}(\gb_i) = \SU(n_i)$, and each $(Z_j,\gb_j)$ is a compact hyper-K\"ahler simply-connected manifold with $\text{Hol}(\gb_j) = \Sp(m_j)$.
\end{enumerate}
It follows that Calabi-Yau $3$-folds can be organized in three distinct families.
\begin{enumerate}[i.]

\item $\Xb$ is a simply-connected Calabi-Yau $3$-fold with $\text{Hol}(\gb) = \SU(3)$.  In this case $\text{Hol}(g) = \SU(3)$ as well, and $\pi_1(X)$ is a finite group.  There are many constructions and a belief that there is a finite number of topological types of such manifolds, but to date no proof of such a finiteness result.  

As an example we can take $\Xb$ to be quintic hypersurface in $\P^4$, where the defining equation is tuned to be invariant under a $G= \Z_5$ action on the projective coordinates with generator
\begin{align}
\label{eq:01234action}
g : [Z_0:Z_1:Z_2:Z_3:Z_4] \mapsto [Z_0:\zeta_5 Z_1:\zeta_5^2 Z_2:\zeta_5^3 Z_3 :\zeta_5^4 Z_4]~,
\end{align}
with $\zeta_k = e^{2\pi i/k}$ a primitive $k$-th root of unity.
The action has $5$ fixed points in $\P^4$, where all but one of the projective coordinates $Z_i$ is zero, but these are missed by a generic $G$--invariant hypersurface.\footnote{One does not have to look far for a representative smooth hypersurface:  the Fermat quintic does the job.}
Since the $G$-action on $X$ is free, we have $\pi_1(X) = G =\Z_5$.

The integral cohomology groups of $X$ are given by~\cite{Batyrev:2005jc}:
\begin{align}
\label{eq:cohomology1and21}
H^0(X) & = \Z~,&
H^2(X) & = \Z \oplus \Z_{5}~,&
H^4(X) & = \Z~,&
H^6(X) & = \Z~,\nonumber\\
H^1(X) & = 0~, &
H^3(X) & = \Z^{2+2\times 21}~, &
H^5(X) & = \Z_5~,
\end{align}
from which we also read off $h^{1,1} = 1$ and $h^{1,2} = 21$; the former is the K\"ahler deformation inherited from the quintic, and the latter are $21$ complex structure deformations, each of which can be represented by a deformation of the $G$--invariant polynomial.  With our definition of the independent torsion subgroups we have $A(X) = \Z_5$, and $B(X) = 0$.  We will return to this example repeatedly in what follows.


\item $\Xb = \text{K3} \times \C$.  Unlike in the first class, $\pi_1(X)$ is infinite.  This time $\text{Hol}^0(g) = \SU(2)$, and $G = \text{Hol}(g)/\text{Hol}^0(g)$ is a finite group.  There is a classification of such quotients, and it is known that there are $8$ deformation families~\cite{Hashimoto:2014oma}.

A familiar example is the Enriques manifold made famous in string theory by~\cite{Ferrara:1995yx} and studied in detail in~\cite{Aspinwall:1995mh}:
\begin{align}
X_{2} =  ((\text{K3}) \times \Ttor^2) /\Z_2~,
\end{align}
where the action of $G=\Z_2$ on a point $(p,z) \in \text{K3}\times\Ttor$ is a combination of the freely-acting Enriques involution $\sigma$ on $\text{K3}$ and reflection in the $\Ttor^2$:
\begin{align}
 (p,z) \mapsto (\sigma(p),-z)~.
\end{align}
The fundamental group is $\pi_1(X_2) = \Z^2 \rtimes \Z_2$, and  $\text{Hol}(g) = \SU(2) \rtimes \Z_2$.

The integral cohomology is~\cite{Aspinwall:1995mh}
\begin{align}
\label{eq:cohomologyEnriques}
H^0(X_2) & = \Z~,&
H^2(X_2) & = \Z^{11} \oplus (\Z_{2})^3~,&
H^4(X_2) & = \Z^{11} \oplus \Z_2~,&
H^6(X_2) & = \Z~,\nonumber\\
H^1(X_2) & = 0~, &
H^3(X_2) & = \Z^{2+2\times 11} \oplus \Z_2~, &
H^5(X_2) & = (\Z_2)^3~,
\end{align}
leading to $h^{1,1}(X_2) = h^{1,2}(X_2) = 11$, $ A(X_2) = (\Z_2)^3$, and $B(X) = \Z_2$.

$X_2$ is self-mirror, and we note that its cohomology is consistent with~(\ref{eq:torsiongroupsmirror})~.  

\item $\Xb = \C^3$.   Now $X$ is a compact flat manifold with holonomy  $\text{Hol}(g) = G$, a finite subgroup of $\SU(3)$ with finite centralizer in $\SU(3)$.  The properties and classification of such manifolds are reviewed in~\cite{Charlap:1986fcm}, and a compact flat manifold necessarily has an infinite fundamental group which cannot contain a finite subgroup~\cite{Auslander:1957hg}.   Such manifolds are classified and come in $6$ deformation families~\cite{Hashimoto:2014oma} (this result is also implicit in the orbifold classification~\cite{Fischer:2012qj}).

As a working example we take the manifold studied in detail in~\cite{Braun:2017oak}:
\begin{align}
X_{\text{3}}= (\Ttor^2\times \Ttor^2\times \Ttor^2)/ (\Z_2\times\Z_2)~,
\end{align}
where the action of the generators of $G = \Z_2 \times \Z_2$ on the three complex coordinates of the elliptic curves is
\begin{align}
g_1 & : (z_0,z_1,z_2) \mapsto \left(z_0 + \ff{1}{2},-z_1,-z_2\right)~, \nonumber\\
g_2 & : (z_0,z_1,z_2) \mapsto \left( -z_0, z_1+\ff{1}{2},-z_2+\ff{1}{2}\right)~.
\end{align}
It is not hard to check that this action is free and projects out $H^1(\Ttor^6,\Z)$, leading to $h^{1,1}(X_{\text{3}}) = h^{1,2}(X_{\text{3}}) = 3$, while $\pi_1(X_{\text{3}})=\Gamma$ is a crystallographic group that fits into the exact sequence
\begin{equation}
\begin{tikzcd}
0 \arrow[r] &\Z^6 \arrow[r] & \Gamma \arrow[r] & G \arrow[r] &0~,
\end{tikzcd}
\end{equation}
where $\Z^6\subset\Gamma$ is generated by elements $k[m_0,m_1,m_2]$ with action
\begin{align}
k[{m}_0, {m}_1,{m}_2]  :(z_0,z_1,z_2) \mapsto (z_0 +m_0,z_1+m_1,z_2+m_2)~,
\end{align}
where $m_a \in \Lambda^a \subset \C$ are lattice vectors of the form $m_{a} = m_{a1} + \tau_a m_{a2}$, and $\tau_a$ specifies the complex structure of the $a$-th elliptic curve.

The integral cohomology is given by
\begin{align}
\label{eq:cohomologyFlat}
H^0(X_3) & = \Z~,&
H^2(X_3) & = \Z^{3} \oplus (\Z_{4})^2 \oplus (\Z_2)^3~,&
H^4(X_3) & = \Z^{3} \oplus (\Z_2)^3~,~~
H^6(X_3)  = \Z~,\nonumber\\
H^1(X_3) & = 0~, &
H^3(X_3) & = \Z^{2 +2\times 3} \oplus (\Z_2)^3~, &
H^5(X_3) & = (\Z_4)^2 \oplus (\Z_2)^3~,
\end{align}
so that $A(X_3) = (\Z_4)^{2} \oplus (\Z_2)^3$, while $B(X_3) = (\Z_2)^3$.

Like $X_2$, $X_3$ is self-mirror, and its cohomology is consistent with~(\ref{eq:torsiongroupsmirror}).

\end{enumerate}
Similar statements can be made for CY $d$-folds with $d>3$, at the price of additional complications and increasing ignorance.  At any rate, the $3$-fold case is the first non-trivial situation, since $d=2$ is more properly considered as a hyper-K\"ahler geometry.

\subsubsection*{A comment on isometries}
The isometry group of $X$ is also strongly constrained.  Suppose $X$ is a compact manifold with a smooth metric $g$.  The group of isometries $\cG_{X,g}$ of $(X,g)$ is a compact Lie group, and its Lie algebra is isomorphic to the Lie algebra of Killing vector fields.  If $\text{Ric}(g) \le 0$, then every Killing vector field is parallel, and the identity component of $\cG_{X,g}$ is isomorphic to a torus of dimension $k \le \dim X$; the orbits foliate $M$ into a parallel family of flat tori.  When $k >0$ $\pi_1(X)$ is necessarily infinite because  the first Betti number is $b_1(X) =k$, and therefore $\{H_1(X,\Z)\}_{\text{free}} \simeq \Z^k$.  The conditions are stronger if $X$ is also K\"ahler~\cite{Ballmann:2006le}.  In that case every Killing vector field is automorphic, meaning it preserves the complex structure.  Now the Lie algebra of Killing vector fields is isomorphic to a complex torus $\Ttor^{2k}$, and $M$ is foliated into a parallel family of flat complex tori; $b_1(X) = 2k$, as it should be on a compact K\"ahler manifold.  
While the existence of a Killing vector necessarily implies that $\pi_1(X)$ is infinite, the converse need not be true.

\subsection{Cohomology, duality, homotopy, and Wall's theorem}
Since all of our manifolds are compact and oriented, it follows that $H_{2d}(X,\Z) = \Z$, and $\{H_{2d-1}(X,\Z)\}_{\text{tor}} = 0$, while their integral homology and cohomology groups are related by Poincar{\'e} duality and the universal coefficients theorem~\cite{Hatcher:2002at}.  The former yields the isomorphism
\begin{align}
\label{eq:Poincaredual}
H^k(X,\Z) \simeq H_{2d-k}(X,\Z)~,
\end{align}
while the latter leads to a relationship between the torsion subgroups that involves Pontryagin duality:
\begin{align}
\label{eq:UniversalCoeff}
\{H^k(X,\Z)\}_{\text{tor}} \simeq \{H_{k-1}(X,\Z)\}_{\text{tor}}^\ast~.
\end{align}
Using these results and our definition, we find that the integral cohomology takes a universal form for all CY $3$-folds:
\begin{align}
\label{eq:cohomologyAll}
H^0(X) & = \Z~,&
H^2(X) & = \Z^{h^{1,1}(X)} \oplus A(X)~,&
H^4(X) & = \Z^{h^{1,1}(X)} \oplus B(X)^\ast,&
H^6(X) & = \Z~,\nonumber\\
H^1(X) & = 0~, &
H^3(X) & = \Z^{2+2\times h^{1,2}(X)} \oplus B(X)~, &
H^5(X_3) & = A(X)^\ast~,
\end{align}
so that the torsion subgroups satisfy
\begin{align}
\{H^{\text{even}}(X)\}_{\text{tor}} & =  A(X) \oplus B(X)^\ast~,&
\{H^{\text{odd}}(X)\}_{\text{tor}} & =  A(X)^\ast \oplus B(X)~.
\end{align}

\subsubsection*{Pontryagin duals}
We recall that the Pontryagin dual $G^\ast$ of a finite abelian group $G$ is simply the set of irreducible representations, with identity element corresponding to the trivial representation.  Equivalently,
\begin{align}
G^\ast = \Hom(G,\GU(1))~.
\end{align}
Any finite abelian group can be brought to the form $G = \Z_{k_1} \oplus Z_{k_2} \oplus \cdots \oplus\Z_{k_N}$, with $k_i$ dividing $k_{i+1}$.  Denoting the elements of $G$ by $\boldsymbol{a} = (a_1,\ldots,a_N)$, with $a_i \in \Z/k_i \Z$, we take $G^\ast$ to have elements $\boldsymbol{b}^\ast = (b^\ast_1,\ldots,b^\ast_N)$, with $b_i \in \Z/k_i \Z$, and the two are related through the non-degenerate pairing
\begin{align}
\label{eq:Pontryaginphase}
e^{2\pi i \la \boldsymbol{b}^\ast, \boldsymbol{a}\ra} &:  G^\ast \times G \to \GU(1)~,  &
\la \boldsymbol{b}^\ast, \boldsymbol{a}\ra = \sum_{i=1}^N \left(\frac{b^\ast_i a_i}{k_i} -  \left[ \frac{b^{\ast}_i a_i}{k_i} \right] \right)~,
\end{align}
where $[x]$ denotes the integer part of $x$.   This pairing has the structure of a discrete Fourier transform, and it is non-degenerate because
\begin{align}
\sum_{\boldsymbol{b}^\ast \in G^\ast} e^{2\pi i \la \boldsymbol{b}^\ast, \boldsymbol{a}\ra} = |G| \delta_{\ba,0}~.
\end{align}

\subsubsection*{Higher homotopy groups}
Recall the relationship between the fundamental group and integral homology:  if $\Gamma = \pi_1(X)$, and $\Gamma_{\text{ab}} = \Gamma/\CO{\Gamma}{\Gamma}$ is its abelianization, then $H_1(X,\Z) = \Gamma_{\text{ab}}$.  When $X$ is simply connected, we have the Hurewicz theorem:  if $\pi_{i}(X)$ is trivial for all $0<i < k$, then $\pi_k(X) = H_k(X,\Z)$.  When $X$ is not simply connected, there is a short exact sequence~\cite{Eilenberg:1945rhh}
\begin{equation}
\label{eq:pi2H2}
\begin{tikzcd}
0 \arrow[r] &\pi_2(X)  \arrow[r] & H_2(X,\Z) \arrow[r] & H_2(\pi_1(X),\GU(1)) \arrow[r] &1~,
\end{tikzcd}
\end{equation}
and this has an important implication for worldsheet instanton sums of the non-linear sigma model path integral described in~(\ref{eq:PIgen}) below.

\subsubsection*{Wall's theorem}
The last classic result we wish to review is due to Wall~\cite{Wall:1966rcd}, which characterizes the diffeomorphism classes of closed, oriented, simply-connected, smooth, spin manifolds $M$ of dimension $6$ and torsion-free homology.  The statement is that the diffeomorphism classes of such $M$ are in bijection with isomorphism classes of systems of invariants that consist of the following data:
\begin{enumerate}
\item two free abelian groups $H = H^2(M,\Z)$ and $G = H^3(M,\Z)$;
\item a symmetric trilinear map $\mu:  H\times H \times H \to \Z$;
\item a homomorphism $p_1:  H \to \Z$~,
\end{enumerate}
subject to, the requirement that for all $x,y \in H$
\begin{align}
\mu(x,x,y) &= \mu(x,y,y) \mod 2~, &
p_1(x) &= 4 \mu(x,x,x) \mod 24~.
\end{align}
In applications to physics there are two aspects of this result that should be borne in mind.  First, it only applies to simply-connected CY $3$-folds without torsion in homology, and does not describe the diffeomorphism classes of the manifolds of interest to us in this work.  While we can always eliminate torsion in $H_1$ by working with the universal cover $\Xb$, there is not an obvious way to eliminate torsion in $H_2$.  Second, we know that while two CY $3$-folds $X_1$ and $X_2$ might belong to distinct diffeomorphism classes, the associated SCFTs may either be isomorphic---for instance $X_1$ and $X_2$ may be a mirror pair, or they may be smoothly connected in the SCFT moduli space:  there may be a path in $\cM_{\text{ac}}$ that connects $\cC[X_1]$ and $\cC[X_2]$ via a finite length curve, even though the geometries have inequivalent triple pairings.  The flop transitions and their generalizations provide an example of the second phenomenon~\cite{Aspinwall:1993nu}.

\subsection{Torsion in K-theory}
Our main interest in this work is in the closed string sector of the string theory compactified on $X$ and the associated worldsheet SCFT defined on compact Riemann surfaces without boundary.  It is not easy to understand the physical significance of a mirror relation such as~(\ref{eq:torsiongroupsmirror}) from this point of view.  But, there is much more to mirror symmetry in the full string theory, where it is a conjectured equivalence of  IIA string theory compactified on $X$ with IIB string theory compactified on $X^\circ$.  In this section we will show that this stronger equivalence implies~(\ref{eq:torsiongroupsmirror}).

Consider the spectrum of D-branes---we think of these as particles in the four-dimensional spacetime---in IIA on $X$, where the closed string background has no topologically non-trivial gerbe for the $B$-field nor any topologically non-trivial Ramond-Ramond fluxes.  In that situation a D-brane configuration is partially classified by an element in the K-theory group $K^0(X)$~\cite{Minasian:1997mm}.\footnote{To our knowledge there is still no complete understanding of how to generalize this statement in  the presence of topologically non-trivial (even if flat) Ramond-Ramond fluxes.  A review of the proposal and its generalization to backgrounds with a non-trivial $B$-field gerbe is given in~\cite{Evslin:2006cj}.}  That is, there is a D-brane for every element in $K^0(X)$, but in general to completely describe the state requires more refined information.\footnote{For example in topological string theory this refined information is encoded by the categorical objects of homological mirror symmetry~\cite{Aspinwall:2009isa}.}     In particular, the torsion subgroup $\{K^0(X)\}_{\text{tor}} \subset K^0(X)$ classifies the torsion branes, which should be stable (though non-BPS) objects in the full string theory.  Similarly, the torsion branes of the IIB string theory on $X^\circ$ are classified by $\{K^1(X^\circ)\}_{\text{tor}}$, and therefore mirror symmetry implies
\begin{align}
\label{eq:Kmirrorstatement}
\{K^0(X)\}_{\text{tor}} \simeq \{K^1(X^\circ)\}_{\text{tor}}~.
\end{align}
In the rest of this section we will show that this statement, when combined with some standard results from the algebraic topology of spin manifolds of dimension $6$ with $H^1(X,\Z) = 0$ and all odd--order generators of $A(X)$, leads to (\ref{eq:torsiongroupsmirror}).

The starting point is the observation that when $X$ is a Calabi-Yau $3$-fold, the Atiyah-Hirzebruch spectral sequence~\cite{Davis:2001at} can be used to relate the torsional cohomology subgroups to the torsional K-theory subgroups~\cite{Brunner:2001eg,Batyrev:2005jc}:
\begin{equation}
\label{eq:KCohomologyTor}
\begin{tikzcd}
0 \arrow[r] &A(X)^\ast  \arrow[r] &  \{K^1(X)\}_{\text{tor}} \arrow[r] & B(X) \arrow[r] &0~,\\
0 \arrow[r] &B(X)^\ast  \arrow[r,"f"] &  \{K^0(X)\}_{\text{tor}} \arrow[r, "c_1"] & A(X) \arrow[r] &0~.
\end{tikzcd}
\end{equation}
We will now see that when $X$ obeys the conditions of the previous paragraph, there are also isomorphisms
\begin{align}
\{K^0(X)\}_{\text{tor}} \simeq A(X) \oplus B(X)^\ast~, &&
\{K^1(X)\}_{\text{tor}} \simeq A(X)^\ast \oplus B(X)~;
\end{align}
in other words, both sequences in~(\ref{eq:KCohomologyTor}) split.

\subsubsection*{Observations on $K^0(X)$}
$K^0(X)$ classifies (virtual) complex vector bundles $X$, and its relation with even cohomology $H^{\text{even}}(X,\Z)$ is given by the set-theoretic map
\begin{equation}
  f:  V \in K(X) \to  (\text{rk}(V),c_1(V),c_2(V),c_3(V)) \in \oplus_{i=0}^{3}H^{2i}(X,\Z)~,
\end{equation}
where $c_i$ is the i-th Chern class, and $\text{rk}(V)$ is the rank of the bundle. Here we want to understand the inverse by characterizing the elements in $H^{\text{even}}(X,\Z)$ that are in the image of this map $f$, or, equivalently, by determining which elements in $H^{\text{even}}(X,\Z)$ could come from an element in K-theory. The result is (see eq 4.27 of \cite{Diaconescu:2000wy}) :
the image of $f$ is given by elements $(r,c_1,c_2,c_3) \in \oplus_{i=0}^{3}H^{2i}(X,\Z)$ that satisfy\footnote{Here $\text{Sq}^2$ is one of the Steenrod cohomology operations; see sect 4.1 of \cite{Diaconescu:2000wy} for an excellent introduction.}
\begin{equation}
\label{K and integral cohomology}
    c_1 \cup c_2 + \text{Sq}^2c_2 + c_3 = 0 \mod 2~.
\end{equation}
We now make the following observations.
\begin{enumerate}
    \item The elements $(r\in \Z,0,0,0) \in H^{\text{even}}(X,\Z)$ can be realized by a rank $r$ trivial virtual vector bundle, and hence belong to the image of $f$.
    \item The elements $(0,\omega \in H^2(X,\Z),0,0)$ can  be realized by a virtual bundle of the form $\cL_{\omega}\ominus \cO$, where $\cO$ a trivial line bundle and $\cL_{\omega}$ is a line bundle with $c_1(\cL_{\omega}) = \omega$, and since the classes satisfy (\ref{K and integral cohomology}), they belong to the image of $f$.
    \item The elements of $(0,0,\upsilon \in H^{4}(X,\Z),0)$ can be realized by a virtual bundle of the form $\cE_{\upsilon} \ominus \cO^{\oplus \text{rk}(\cE_{\upsilon})}$, where $\cE_{\upsilon}$ is a vector bundle with $c_1(\cE_{\upsilon}) = 0$, $c_2(\cE_{\upsilon}) = \upsilon$, and $c_3(\cE_{\upsilon}) = 0$.  Such a class is in the image of $f$ because for $X$ spin, $\text{Sq}^2\upsilon = w_2(X)\cup \upsilon =0$.  
    \item The element $(0,0,0, \rho \in H^6(X,\Z))$ can be realized by a virtual vector bundle if and only if  $\rho = 0\mod2$; e.g. $c_3(\cO_p) =2$ for a skyscraper sheaf on CY threefolds.
\end{enumerate}
Putting together these observations, we see that the map $f$ is almost surjective, with the only subtlety being the restriction $\rho = 0 \mod 2$.  However, when we restrict to the torsion subgroups, this subtlety disappears because $\{H^6(X,\Z)\}_{\text{tor}} = 0$.  For the same reason the constraint~(\ref{K and integral cohomology}) becomes trivial, so that restricting the map $f$ to the torsion subgroups we obtain  the surjective set-theoretic map 
\begin{align}
f_0: \{K^0(X)\}_{\text{tor}} \to A(X) \oplus B(X)^\ast~.
\end{align} 

\subsubsection*{Splitting the sequences}
The preceding observations do not by imply that the sequence for $\{K^0(X)\}_{\text{tor}}$ splits.  The desired splitting is equivalent to the existence of a surjective group homomorphism $h : \{K^0(X)\}_{\text{tor}} \to H^4(X,\Z)$.  Taking rational coefficients the Chern character provides the group isomorphism $\ch  : K^0(X) \otimes_{\Z} \Q \to H^{\text{even}}(X,\Q)$~\cite{Hatcher:2009vb}.  Restricting this to the second Chern character we find the surjective map $\ch_2:  K^0(X)\otimes_{\Z} \Q \to H^4(X,\Q)$, with image
\begin{align}
\ch_2(E) = \ff{1}{2} c_1(E) \cup c_1(E) - c_2(E)~.
\end{align}
Generally this does not lead to a well-defined map on integer classes because $c_1(E) \cup c_1(E)$ need not be divisible by $2$.  
However, suppose now that $H^2(X,\Z)_{\text{tor}}$ has no elements of even order.\footnote{Every $x \in H^2(X,\Z)_{\text{tor}}$ has an order --- the smallest positive integer $p$ such that $px = 0$.}  In this case the order $p$ of any element $x \in H^2(X,\Z)_{\text{tor}}$ is odd, and we can write $x = 2 \ff{(p+1)}{2} x$, which shows that $\ff{1}{2} x \cup x$ is a well-defined class in $H^4(X,\Z)_{\text{tor}}$.  Applying this observation to $x = c_1(E)$, we now conclude that with our assumption on $A(X) = H^2(X,\Z)_{\text{tor}}$ there is a surjective group homomorphism
\begin{align}
g&:   K^0(X)_{\text{tor}} \to B(X)^\ast~, &
g(E) =  \ff{1}{2} c_1(E) \cup c_1(E) - c_2(E)~.
\end{align}
Moreover, by observation 3 above this map remains surjective even when restricted to $\im f = \ker c_1$, and that means $\phi= g \circ f$ is a surjective finite group endomorphism  $\phi : B(X)^\ast \to B(X)^\ast$ and hence an isomorphism, so that we obtain the split sequence
\begin{equation}
\label{eq:K0split}
\begin{tikzcd}
0 \arrow[r] &B(X)^\ast  \arrow[r,"f \phi^{-1}"] &  \{K^0(X)\}_{\text{tor}} \arrow[r, "c_1"] \arrow[bend left= 30] {l}{g} & A(X) \arrow[r] &0~.
\end{tikzcd}
\end{equation}
The sequence for $K^1(X)_{\text{tor}}$ splits as well.  This follow from the perfect pairing
(see section 3.1 of ~\cite{Moore:1999gb}):
\begin{align}
\{K^0(X)\}_{\text{tor}} \times \{K^1(X)\}_{\text{tor}} \to \GU(1)~,
\end{align}
which implies $\{K^0(X)\}_{\text{tor}}$ is the Pontryagin dual to $K^1(X)$, and therefore
\begin{align}
\label{eq:K1split}
\{K^1(X)\}_{\text{tor}} = A(X)^\ast \oplus B(X)~.
\end{align}
We have not succeeded in generalizing this proof to the situation when $A(X)$ has elements of even order, but we have been able to show that for the examples $X_2$ and $X_3$ from section~\ref{ss:holofun} the splitting~(\ref{eq:K0split}) still holds, even though every element of $A(X)$ has an even order.   Thus, it may be possible to generalize the argument to establish~(\ref{eq:K0split}) for CY threefolds.

Having shown the split, we note that our assumptions and mirror symmetry imply that $A(X^\circ)^\ast \subset K^1(X^\circ)_{\text{tor}}$ and $B(X^\circ)^\ast \subset K^0(X^\circ)_{\text{tor}}$ also only have elements of odd degree, and thus the sequences mirror to~(\ref{eq:KCohomologyTor}) split as well.  Thus, mirror symmetry and absence of even order torsion cocycles in $H^\ast(X,\Z)$  imply~(\ref{eq:torsiongroupsmirror}).

\section{Orbifolds, covers, and mirrors} \label{s:orbifoldcovers}
In this section we examine the role of $A(X) = \{H^2(X,\Z)\}_{\text{tor}}$ in the SCFT $\cC[X]$.  We will show that it has a physical interpretation as the quantum symmetry of an orbifold SCFT.

\subsection{Intuition from the non-linear sigma model}
Let $X$ be a simply-connected Calabi-Yau $3$-fold equipped with a metric $g$ and flat gerbe $\beta \in H^2(X,\GU(1))$.  When $X$ is smooth, we can hope to define the SCFT $\cC[X]$, at least in a large volume limit, by a NLSM of maps from the worldsheet $\Sigma \to X$.  At string tree-level we take $\Sigma = \P^1$.  The NLSM path integral then breaks up into topological sectors labeled by $\pi_2(X)$, and since $X$ is simply-connected, it follows from~(\ref{eq:pi2H2}) that we can equivalently label the topological sectors by classes in $H_2(X,\Z)$.  Formally, the NLSM partition function is given by
\begin{align}
\label{eq:PIgen}
Z[\Sigma;X,g,\beta] &= \sum_{C \in H_2(X,\Z)} \int_{[f(\Sigma)] = C} [\text{D} f \text{D}\psi] e^{-S[X,g; f,\psi]}
\nonumber\\
&\qquad \qquad \qquad\qquad\qquad \times \exp\left[ 2\pi \ii \left( \int_{C_{\text{f}}} \beta_{\text{f}} + \la C_{\text{t}},\beta_{\text{t}}\ra\right) \right]~.
\end{align}
Here $\psi$ refers to the worldsheet fermion degrees of freedom, while $S$ is the usual $(2,2)$ supersymmetric action for a non-linear sigma model with K\"ahler target space.\footnote{We will not need explicit details of this action, but they can be found in many reviews and textbooks, e.g.~\cite{Deligne:1999qp,Hori:2003ds}.  The references also review the evidence for the existence of these quantum field theories.}  Our main interest is the phase factor in the second line that encodes the dependence of the theory on the gerbe.  
To give an explicit form of the gerbe phase factor we chose a basis for
\begin{align}
H_2(X,\Z) = \underbrace{\Z^{h^{1,1}(X)}}_{ \ni C_{\text{f}}} \oplus \underbrace{B(X)^\ast}_{\ni C_{\text{t}}}~,
\end{align}
and a dual basis for $H^2(X,\GU(1))$, so that $\beta$ is represented by the pair $\beta = (\beta_{\text{f}},\beta_{\text{t}})$, with $\beta_{\text{f}} \in H^2(X,\R)/H^2(X,\Z)$, while $\beta_{\text{t}} \in B(X)$.  The first term in the phase is the familiar worldsheet coupling to a closed $B$-field, while the second one uses the Pontryagin pairing to encode the additional phase that depends on the torsion cycle $C_{\text{t}} \in H_2(X,\Z)$ and on $\beta_{\text{t}}$.\footnote{This presentation that keeps track of the torsion cycles goes back to~\cite{Aspinwall:1995rb}.  Recently it played a role in the study and definition of refined Gopakumar-Vafa invariants and their relations to exotic gauged linear sigma model phases~\cite{Katz:2022lyl,Katz:2023zan}.  More generally such couplings should be understood in the language of differential cohomology---see, for example,~\cite{Freed:2006yc}.}

Giving a rigorous definition of this path integral and showing that the resulting theory is conformal remains an outstanding challenge.  However, there is substantial circumstantial evidence that the path integral exists, and it leads to families of SCFTs of central charge $c=\cb=9$ labeled by a choice of complexified K\"ahler class $\beta_{\text{f}} + i J \in H^2(X,\C)/H^2(X,\Z)$ and complex structure on $X$.  When $X$ is smooth, the NLSM becomes weakly coupled as $J$ is taken to be deep in the interior of the K\"ahler cone of $X$.  This is the large volume limit.

Although so far we have only discussed the path integral with worldsheet $\Sigma = \P^1$, for our closed string applications we need to define it more generally on any closed Riemann surface $\Sigma_g$.  This requires a choice of spin structure for the fermions, and by summing appropriate combinations of these, we can obtain modular invariant partition functions.  Of particular importance is the theory on $\Sigma_1 = \Ttor^2$, where a modular-invariant partition function can be obtained by taking the usual diagonal modular invariant with a non-chiral GSO projection in the NS-NS and R-R sectors~\cite{Seiberg:1986by,Ginsparg:1988ui}. 

\subsection{A NLSM orbifold}
If $(X,g)$ has a non-trivial isometry group $\cG_{X,g}$, then by the results reviewed above $\cG_{X,g}$ is necessarily finite.  Let  $\Gamma \subset \cG_{X,g}$ be a freely-acting subgroup of K\"ahler isometries, i.e. $\Gamma$ preserves the metric and the complex structure on $X$.  Although initially just defined on the bosonic degrees of freedom, we can extend the action of $\Gamma$ to the left- and right-moving fermions in a symmetric fashion, and if we choose the gerbe to be trivial, i.e. $\beta_f = 0$ and $\beta_{\text{t}} = 0$, then we expect that $\Gamma$ will be a symmetry of the NLSM, and therefore of the SCFT $\cC[X]$.  We can then attempt to construct the orbifold SCFT $\cC[X]/\Gamma$ by introducing twisted sectors labeled by the conjugacy classes of $\Gamma$ and projecting to $\Gamma$--invariant states.  In doing so, we encounter a number of potential subtleties:
\begin{enumerate}
\item $\Gamma$ may have `t Hooft anomalies that prevent us from gauging it, but since our action is defined symmetrically on the left- and right-moving degrees of freedom, we do not expect to encounter this difficulty: there should be a regularization scheme that explicitly preserves the $\Gamma$ symmetry.
\item In order to embed this SCFT construction in a supersymmetric string theory we need $\Gamma$ to preserve (2,2) supersymmetry and integrality of the $\GUL\times\GUR$ charges in the NS-NS sector.  Because $\Gamma$ preserves the K\"ahler structure, the preservation of (2,2) supersymmetry is guaranteed, but the integrality of R-charges may indeed be violated.  For example, this happens if $X$ is a K3 surface and $\Gamma = \Z_2$ is the group generated by the Enriques involution.  However, when $X$ is a CY $3$-fold every freely-acting K\"ahler isometry preserves the holomorphic $3$-form on $X$, and therefore we expect the orbifold to maintain the requisite charge integrality.  Further discussion of this point is given in~\cite{Cheng:2023owv}.

\item Although orbifold CFTs can be constructed quite generally at the abstract level~\cite{Brunner:2013xna,Bhardwaj:2017xup}, an explicit description of the twisted sectors is not straightforward.  However, with a path integral presentation, the twisted sectors amount to specifying the boundary conditions on the NLSM fields, as was originally discussed in~\cite{Dixon:1986jc,Dixon:1987bg,Ginsparg:1988ui}~.  

\item There is an ambiguity in the action of $\Gamma$ on the twisted Hilbert spaces which can lead to several inequivalent orbifold theories.  This ambiguity is encoded in the choice of discrete torsion~\cite{Vafa:1986wx}\footnote{In modern terminology this can be understood as data needed to determine the ``orbifolding defect'' in the language of~\cite{Brunner:2014lua} or including an additional TFT in the construction~\cite{Gaiotto:2014kfa,Bhardwaj:2017xup}.}
 --- a phase factor in the orbifold partition function defined on the worldsheet $\Sigma_1 = \Ttor^2$, a torus with complex structure parameter $\boldsymbol{\tau}$ and modular parameter $\bq = e^{2\pi \ii\boldsymbol{\tau}}$.   Let $\Theta$ denote the set of conjugacy classes $\theta $ of $\Gamma$, and for each $\theta \in\Theta$ let $\cN(\theta)\subset \Gamma$ be the stabilizer subgroup.  For each $\theta\in \Theta$ we have a twisted Hilbert space $\cH_{\theta}$, and for each $\gamma \in \cN(\theta)$ a unitary operator $U_{\theta}(\gamma)$ that represents the action of $\cN(\theta)$ on $\cH_{\theta}$.  The orbifold partition function takes the form
\begin{align}
\label{eq:orbpartitionfunc}
Z_{\Gamma}(\boldsymbol{\tau},\overline{\boldsymbol{\tau}}) & =  \sum_{\theta \in\Theta} \frac{1}{|\cN(\theta)|} \sum_{\gamma\in\cN(\theta)} \varepsilon(\gamma,\gamma_{\theta})\Tr_{\cH_{\theta}}  \left(  U_{\theta}(\gamma) \bq^{L_0-c/24} \overline{\bq}^{\Lb_0-\cb/24}\right)~,
\end{align}
where $L_0$ and $\Lb_0$ are the usual Virasoro generators, $\gamma_\theta$ is any group element in the conjugacy class $\theta$, and the phase $\varepsilon(\gamma_1,\gamma_2) \in H^2(\Gamma,\GU(1))$ is the discrete torsion of~\cite{Vafa:1986wx}, which must satisfy
\begin{align}
\label{eq:discretetorsion}
\varepsilon(\gamma_1\gamma_2,\gamma_3) & = \varepsilon(\gamma_1,\gamma_3) \varepsilon(\gamma_2,\gamma_3)~,&
\varepsilon(\gamma_1,\gamma_2) \varepsilon(\gamma_2,\gamma_1) & =1~,&
\varepsilon(\gamma_1,\gamma_1) & = 1 
\end{align}
for all $\gamma_{1},\gamma_{2},\gamma_{3} \in\Gamma$.  We see  that $\varepsilon(\gamma,\gamma_\theta)$ only depends on the conjugacy class of $\gamma_\theta$. 

The expression simplifies considerably when $\Gamma = G$ is abelian, and since that will be our main focus, we give the expression now, using our notation for abelian groups from above~(\ref{eq:Pontryaginphase}):
\begin{align}
\label{eq:abelianorbpartitionfunc}
Z_{G}(\boldsymbol{\tau},\overline{\boldsymbol{\tau}}) & =  \frac{1}{|G|} \sum_{\ba,\bb \in G}  \varepsilon(\bb,\ba) Z^{\bb}_{\ba}(\boldsymbol{\tau},\overline{\boldsymbol{\tau}})~,\nonumber\\
Z^{\bb}_{\ba} (\boldsymbol{\tau},\overline{\boldsymbol{\tau}}) &=
\Tr_{\cH_{\ba}}  \left( U_{\ba}(\bb) \bq^{L_0-c/24} \overline{\bq}^{\Lb_0-\cb/24}\right)~.
\end{align}

\item In the context of the orbifold of the NLSM for a target space $X$ by a group of K\"ahler isometries $\Gamma$, we should also discuss the role of the flat gerbe.  The first point is that if the gerbe is non-zero, then the $\Gamma$ action should preserve the gerbe structure in order to be a symmetry of the NLSM.  Moreover, if the gerbe is topologically non-trivial, i.e. $\beta_{\text{t}} \neq 0$, a more careful discussion of potential anomalies in gauging $\Gamma$ is necessary.   When $\beta_{\text{t}} = 0$ a sufficient condition for $\Gamma$ to be a symmetry group of the NLSM is that for every $\gamma \in \Gamma$, with corresponding diffeomorphism $\varphi_{\gamma} : X\to X$, the gerbe connection $\beta_{\text{f}}$ represented by a closed $2$-form satisfies
\begin{align}
\varphi_{\gamma}^\ast(\beta_{\text{f}}) - \beta_{\text{f}} \in H^2(X,\Z)~.
\end{align}
The second point is that in taking the orbifold we have the possibility of introducing a $\Gamma$-equivariant gerbe structure~\cite{Sharpe:2000ki,Sharpe:2003cs,Cheng:2022nso} to modify the orbifold theory.  The modifications include the introduction of discrete torsion, but they also go beyond it: for example, the orbifold may admit a non-trivial equivariant gerbe even when $H^2(\Gamma,\GU(1))$ is trivial. 

\end{enumerate}
Suppose now that we have a NLSM for a simply-connected target space $\Xb$ with $\beta_{\text{t}} =0$ and a finite freely-acting symmetry $\Gamma$ just as above.  Then, starting with the SCFT $\cC[\Xb]$ with the global symmetry $\Gamma$, we construct the orbifold $\cC[\Xb]/(\Gamma,\varepsilon)$.  When $\Xb$ is large and smooth, the resulting theory has a
geometric interpretation as $\cC[X,\beta_{\text{t}}]$, where $X = \Xb/\Gamma$ is the smooth geometric quotient, and $\beta_{\text{t}}$ is the topologically non-trivial gerbe on $X$ that encodes the choice of discrete torsion $\varepsilon$.

The orbifold SCFT is naturally equipped with a global symmetry---the so-called quantum symmetry~\cite{Vafa:1989ih}, which acts by phases on the twisted sectors.  This group is $\Gamma_{\text{q}} \simeq (\Gamma_{\text{ab}})^\ast$, where $\Gamma_{\text{ab}} = \Gamma/\CO{\Gamma}{\Gamma}$, and comparing to our geometric discussion from above, we recognize its geometric significance.  Since $\pi_1(X) \simeq \Gamma$, it follows that $H_1(X,\Z) \simeq \Gamma_{\text{ab}}$, and the torsion subgroup $A(X) \subset H^2(X,\Z)$ is the quantum symmetry
\begin{align}
A(X) = \Gamma_{\text{q}}~.
\end{align}
When $\Gamma = G$ is abelian, in which case $G_{\text{q}} = G^\ast$, the action on the twisted sectors is expressed via the Pontryagin pairing of~(\ref{eq:Pontryaginphase}): for every $\bb^\ast \in G^\ast$ we have the action
\begin{align}
\bb^\ast : |\psi\ra \to e^{2\pi i \la \bb^\ast,\ba\ra} |\psi\ra~,\qquad\text{for all}~~|\psi\ra \in \cH_{\ba}~.
\end{align}

The quantum symmetry puts the orbifold theory on the same footing as the parent theory:  we can recover $\cC[\Xb]$ as $\cC[\Xb] = (\cC[\Xb]/G)/G_{\text{q}}$ by gauging the quantum symmetry $G_{\text{q}}$ of the orbifold $\cC[\Xb]/G$.  The equivalence can be seen at the level of the partition function as a consequence of a discrete Fourier transform.  Starting with~(\ref{eq:abelianorbpartitionfunc}), the
partition function for $(\cC[\Xb]/G)/G_{\text{q}}$ is 
\begin{align}
Z' = \frac{1}{|G_q|}\sum_{\ba^\ast,\bb^\ast \in G_{\text{q}}}   \frac{1}{|G|}\sum_{\ba,\bb \in G}e^{2\pi \ii ( \la \ba^\ast,\ba\ra + \la \bb^\ast,\bb\ra)} 
\varepsilon(\bb,\ba)Z^{\bb}_{\ba}  = Z~,
\end{align}
and the last equality follows by exchanging the order of summation.  Note that we recover the original partition function for any choice of discrete torsion in the original quotient because $\varepsilon(0,0) = 1$.  It is also possible to recover the original theory for a general non-abelian quotient by $\Gamma$ by gauging a suitable categorical symmetry in the orbifold~\cite{Bhardwaj:2017xup}.

\subsubsection*{Moduli and twisted strings in CY orbifolds}
Our discussion so far has not touched on aspects of the quotient particular to theories with (2,2) superconformal invariance preserved by the gauging.   Because the action is free, the orbifold twisted sectors do not introduce new marginal deformations, and instead the moduli space $\cM_{\text{cc}}(X) \times \cM_{\text{ac}}(X)$ is obtained by restricting $\cM_{\text{cc}}(\Xb) \times \cM_{\text{ac}}(\Xb)$ to the $\Gamma$--invariant locus.  Moreover, the correlation functions on $\Sigma = \P^1$ of the $\Gamma$--invariant operators from the untwisted sector remain unmodified.\footnote{This ``orbifold inheritance principle'' is reviewed in~\cite{Polchinski:1998rq}.}  Thus, the chiral ring structure is also inherited from the parent theory.\footnote{This has the implication that ``chiral rings do not suffice'' to distinguish (2,2) SCFTs~\cite{Aspinwall:1994uj}.  When $H_2(\pi_1(X),\GU(1)) \neq 1$ the sequence~(\ref{eq:pi2H2}) implies that there are homology cycles that cannot be represented by spheres embedded in $X$.  And thus, even though the chiral ring of $X$ is just inherited by restricting that of $\Xb$, there can be higher genus correlation functions that distinguish the $\cC[X]$ from the parent theory.}
We also note that since all marginal deformations are uncharged with respect to $\Gamma_{\text{q}}$, the quantum symmetry remains unbroken for all points in the moduli space.

Since $\Gamma$ acts freely, we also have the relationship between the dimensions of the spaces given by the Euler number, which satisfies
\begin{align}
\chi(\Xb) = 2(h^{1,1}(\Xb) - h^{1,2}(\Xb)) = |\Gamma| \chi(X)  = 2 |\Gamma| (h^{1,1}(X) - h^{1,2}(X))~.
\end{align}
This implies that such orbifolds can lead to CY $3$-folds with small Hodge numbers, and ever since~\cite{Candelas:1985en} there has been a substantial effort to characterize these free quotients.  Some results and classifications are given in~\cite{Braun:2010vc,Gray:2021kax,Candelas:2016fdy,Braun:2017juz}.

While the twisted sectors do not contain additional marginal deformations, they do contain non-BPS states that describe strings that wrap the non-contractible cycles of $\pi_1(X)$.  These ``winding modes'' are charged under the quantum symmetry of the orbifold, and of course they are very massive when $X$ is large, with a mass proportional to the $R/\ell_s^2$, where $R$ is the length of the cycle and $\ell_s$ is the string length.

So, the upshot is that we understand the significance of the torsion subgroup $A(X) \in H^2(X,\Z)$.  Its presence indicates that the SCFT $\cC[X]$ has a global symmetry  at every point in its moduli space, and gauging this symmetry recovers the SCFT based on the universal cover $\Xb$.  Note that the symmetry action is intrinsically stringy because only winding strings are charged with respect to it.

\subsection{Mirror symmetry}
Having described the role of the group $A(X)$ in $\cC[X]$, we now turn to its interpretation in the mirror description, and we begin with an example:  a $\Z_5$ quotient of the quintic described in section~\ref{ss:holofun}, and its mirror.

\subsubsection*{A glimpse of toric geometry and linear sigma models}
Let $\Xb$ be a generic hypersurface in $V = \P^4$.  Its mirror $\Xb^{\circ}$ is a quintic hypersurface in a toric variety $V^\circ$ obtained by a resolution of singularities in $\P^4/(\Z_5)^3$.   The relationship between the SCFTs can be understood in terms of the Batyrev construction~\cite{Batyrev:1994hm} nicely reviewed in~\cite{Cox:2000vi}, which specifies the data for a pair of two-dimensional gauged linear sigma models (GLSMs)~\cite{Witten:1993yc,Morrison:1994fr} that flow at low energy to isomorphic SCFTs.\footnote{We resist the temptation to launch into a detailed review of toric geometry.  Additional details and references can be found in~\cite{Melnikov:2019tpl}, whose notation we follow closely in what follows.  A comprehensive and comprehensible treatment of toric geometry is given in~\cite{Cox:2011tv}. }  After recalling the key aspects of that construction, we will see that it also allows us to understand the quotient and its mirror.

The combinatorial data for the gauge theory is encoded in a dual pair of reflexive polytopes $\Delta^\circ \subset N_{\R}$ and $\Delta \subset M_{\R}$, where $N$ and $M$ are dual lattices, each isomorphic to $\Z^d$ (for our $3$-fold examples $d=4$) and embedded in a Euclidean space $N_{\R} = N \otimes_{\Z} \R$ and $M_{\R} = M\otimes_{\Z} \R$.   For our quintic example these polytopes have vertices given by the columns of $\text{Vert}_{\Delta^\circ}$ and the rows of $\text{Vert}_{\Delta}$, with
\begin{align}
\label{eq:quinticpolytopes}
\text{Vert}_{\Delta^\circ} & = 
\begin{pmatrix} \rho_0 & \rho_1 &\cdots & \rho_4\end{pmatrix}=
\begin{pmatrix}
-1& 1&0&0&0 \\
-1& 0&1&0&0 \\
-1& 0&0&1&0\\
-1& 0&0&0&1 
\end{pmatrix}~,
&
\text{Vert}_{\Delta} & = 
\begin{pmatrix}
v_0 \\ v_1 \\ v_2 \\ v_3 \\ v_4 
\end{pmatrix}
=
\begin{pmatrix}
-1 & -1 & -1 & -1 \\
~4 & -1 & -1 & -1 \\
-1 & ~4 & -1 & -1 \\
-1 & -1 & ~4 &-1 \\
-1 & -1 & -1 & ~4
\end{pmatrix}~.
\end{align}
$\Delta^\circ$ has no additional non-zero lattice points, while $\Delta$ has a total of $126$ lattice points.

We obtain a fan $\Sigma_V$ for a toric variety $V$ by taking the cones over the faces of $\Delta^\circ$ and choosing a maximal projective subdivision.  Such a subdivision requires that the one-dimensional cones, i.e. the rays in $\Sigma_V$, include every non-zero lattice point in $\Delta^\circ$.  We denote those lattice points by $\rho \in (\Delta^\circ \cap N) \setminus \{0\}$.  The toric variety $V$ can be presented as a quotient
\begin{align}
\label{eq:toricquotient}
V = \frac{ \C^n \setminus F} {\cG_{\C}}~,
\end{align}
where $n = |(\Delta^\circ \cap N) |-1$.  The exceptional set $F$ is fixed by the choice of maximal projective subdivision, and the quotient group $\cG_{\C}$ is determined by the short exact sequence
\begin{equation}
\label{eq:gaugegroupsequence}
\begin{tikzcd}
1\ar[r] & \cG_{\C} \ar[r]  &(\C^\ast)^n \ar[r,"R"] & T_N \ar[r] & 1~,
\end{tikzcd}
\end{equation}
where $T_N = N\otimes_{\Z} \C^\ast \simeq \Hom_{\Z} (M,\C^\ast)$ can be identified with the algebraic torus contained as a dense subset in $V$, and the map $R$ is determined by the lattice points $\rho$.  It follows that $\cG_{\C}$ is the complexification of an abelian group $\cG \simeq \GU(1)^{n-d} \times K$, where $K$ is a finite abelian group.  

The linear sigma model is obtained as follows.  There are $n$ (2,2) chiral superfields $Z_\rho$, and these are minimally coupled to $n-d$ abelian vector multiplets, with charges determined by $\cG$.  We introduce an additional chiral superfield $\Phi_0$ with gauge charges chosen such that $\Phi_0 \prod_{\rho} Z_\rho$ is gauge-invariant, and we set the chiral superpotential to be
\begin{align}
W &= \Phi_0 P(Z)~, &
P(Z) & =  \sum_{m \in \Delta \cap M} \alpha_m \prod_\rho Z_\rho^{\la m,\rho\ra + 1}~.
\end{align}
The $\alpha_m$ are generic complex coefficients, and $\la m,\rho\ra$ denotes the pairing between the $M$ and $N$ lattices.  The remaining term in the Lagrangian is the twisted chiral superpotential, which is linear in the gauge-field strengths of the vector multiplets and introduces $n-d$  Fayet-Iliopoulos parameters $r_\alpha$ complexified by $n-d$ $\theta$-angles $\theta_\alpha$.  The theory has a non-anomalous $\GUL\times\GUR$ R-symmetry, and for a suitable choice of the $r_\alpha$, it is believed to flow at low energy to the same fixed point as the NLSM based on the CY hypersurface $\Xb$ defined by the locus
\begin{align}
\Xb = \{P(Z)  = 0 \} \subset V~.
\end{align}

Mirror symmetry is the assertion that an isomorphic SCFT arises as the low energy limit of the linear sigma model obtained by exchanging the roles of $\Delta^\circ$ and $\Delta$.  In the mirror theory the non-zero lattice points of $\Delta$ determine the toric variety $V^\circ$, while the lattice points of $\Delta^\circ$ determine the monomials used to construct the mirror hypersurface $\Xb^\circ = \{ P^\circ = 0\} \subset V^\circ$.  The construction naturally incorporates the monomial--divisor mirror map~\cite{Aspinwall:1993rj}, which identifies a subset of complexified K\"ahler deformations---the ``toric'' deformations obtained by varying the complexified FI parameters of the linear sigma model, with a subset of the complex structure deformations of the mirror---the ``polynomial'' deformations obtained by varying the coefficients of monomials in the chiral superpotential.\footnote{A recent discussion of these deformations and the fate of the remaining non-toric/non-polynomial ones can be found in~\cite{Adams:2023imc}.}

\subsubsection*{An example of a quotient and its mirror}
Having glanced at the essential features of the Batyrev mirror symmetry, we proceed to study our example.  Starting with the mirror pair $\Xb$, $\Xb^\circ$, we observe that we can tune the coefficients $\alpha_m$ for the quintic polynomial in $\P^4$ so that the linear sigma model (and the geometry) acquires an additional $G = \Z_5$ symmetry, which acts by phases as in~(\ref{eq:01234action}).   The action on a monomial corresponding to lattice point $m \in \Delta$ is to multiply it by a phase $\zeta_5^{\la m,\rho_g\ra}$, where $\rho_g = \rho_1 + 2\rho_2 + 3\rho_3 + 4\rho_4$.  We say $m$ is a $G$--invariant lattice point if $\la m,\rho_g \ra = 0 \mod 5$, and it is not hard to see that there are $26$ of these.  They consist of the origin, the vertices $v_0,\ldots,v_4$, as well points contained in the relative interior of dimension $2$ faces $F_{abc}$
\begin{align}
F_{abc} = \{ t_a v_a + t_b v_b + t_c v_c~~|~~ t_{a,b,c}\ge 0~, t_a + t_b +t_c = 1\}~,
\end{align}
with each $F_{abc}$ containing precisely $2$ invariant lattice points in the interior.  For example, we have the face $F_{012}$
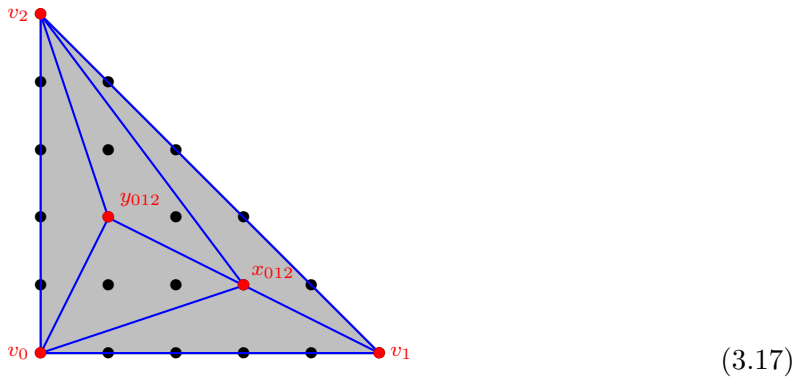
\begin{align}
\label{eq:Facepicture}
\begin{tikzpicture}[scale=0.9]
\filldraw[black!25] (-1,4)--(-1,-1)--(4,-1)--cycle;
\foreach \y in {0,...,5}
 \foreach \x in {\y,...,5}
   \draw (4-\x,\y-1) node {$\bullet$};
\draw[blue,thick]  (4,-1)--(-1,4)--(-1,-1)--cycle;
\draw[blue,thick] (4,-1) -- (0,1);
\draw[blue,thick] (0,1) -- (-1,-1);
\draw[blue,thick] (0,1) -- (-1,4);
\draw[blue,thick] (2,0) -- (-1,-1);
\draw[blue,thick] (2,0) -- (-1,4);
\draw[red]  (-1,-1) node {$\bullet$} node[left] {\scriptsize $v_0$};
\draw[red]  (-1,4) node {$\bullet$} node[left] {\scriptsize $v_2$};
\draw[red]  (4,-1) node {$\bullet$} node[ right] {\scriptsize $v_1$};
\draw[red]  (2,0) node {$\bullet$} node[above right =-2pt] {\scriptsize $x_{012}$};
\draw[red]  (0,1) node {$\bullet$} node[above right] {\scriptsize $y_{012}$};
\end{tikzpicture}
\end{align}
where we marked in red the invariant lattice points, and the blue indicates a triangulation that we will discuss presently.

Once we set $\alpha_m = 0$ for all non-invariant lattice points and keep the remaining $\alpha_m$ generic, we obtain a family of quintic hypersurfaces $\Xb \subset \P^4$, each with a freely-acting isometry $G$.  The moduli space of these theories has the form $\cM_{\text{ac}}(\Xb) \times \cM_{\text{cc}}(\Xb)_{G}$, where the subscript indicates that we restrict to the $G$--invariant locus in the quintic moduli space.  Note that this places no restriction on the complexified K\"ahler parameter. Taking the quotient, we obtain our example $X$ with $h^{1,1}(X) = 1$, $h^{1,2}(X) = 21$, $A(X) = G^\ast$, and $B(X) = 0$.  By the arguments in the previous section the SCFT moduli space is given by $\cM_{\text{ac}}(X) \times \cM_{\text{cc}}(X) = \cM_{\text{ac}}(\overline{X})\times \cM_{\text{cc}}(\Xb)_{G}$.

To understand the mirror of this procedure, we follow the same steps.  First, mirror symmetry implies that there is a locus in the complexified K\"ahler moduli space of $\Xb^\circ$, where the SCFT acquires an additional global symmetry $G^\circ \simeq G$, and we have the mirror isomorphism $\cM_{\text{ac}}(\Xb^\circ)_{G^\circ} \simeq \cM_{\text{cc}}(\Xb)_G$.  The variety $\Xb$ is necessarily singular, as can be seen from~(\ref{eq:Facepicture}).  This follows because now that we interpret $\Delta$ as the polytope that leads to the fan $\Sigma_{V^\circ}$, we see that this fan includes many full-dimensional cones that do not generate the full lattice:   each non-zero lattice point in $\Delta^\circ$ corresponds to a toric divisor, and by restricting to the invariant lattice points we have blown down all but the ones associated to the red points in~(\ref{eq:Facepicture}).  Nevertheless, mirror symmetry implies that the SCFT remains smooth, and the orbifold by $G^\circ$ leads to the mirror $X^\circ$.

As we will see, the action of $G^\circ$ does not have a simple geometric interpretation.  In a sense this is perhaps to be expected, since the action is defined on $\Xb^\circ$ --- a singular space on the $G^\circ$--invariant locus in $\cM_{\text{ac}}(\Xb)$.  However, we will now show that it can be understood as a quantum symmetry of an orbifold that does have a simple geometric interpretation.  To see this, we will use toric morphisms to describe the mirror dual of the quantum symmetry $(G_{\text{q}})^\circ$.

\subsubsection*{An aside on toric morphisms}
We follow the presentation in chapter 3 of~\cite{Cox:2011tv} but express the results in homogeneous coordinates which are more appropriate for our discussion. 

Let $V$ be a toric variety determined by a fan $\Sigma \subset N$, but suppose that the generators of $\Sigma$ in fact belong to a finite index sub-lattice $N' \subset N$, with $|N/N'| >1$. The embedding $L : N'\to N$ determined by writing the generators of $N'$ in terms of those of $N$ gives a map $\Lambda : T_{N'} \to T_N$, and this leads to a relationship between the toric varieties $V = V_{\Sigma \subset N}$ and $V' = V_{\Sigma \subset N'}$.  Each of these has a presentation as a holomorphic quotient of the form~(\ref{eq:toricquotient}), and the two quotients only differ in the group actions, which are related by
\begin{equation}
\begin{tikzcd}
1\ar[r] & \cG'_{\C} \ar[r]  &(\C^\ast)^n \ar[r,"R'"] ,\ar[d,"\text{id}"] & T_{N'} \ar[r]  \ar[d,"\Lambda"]& 1~\\
1\ar[r] & \cG_{\C} \ar[r]  &(\C^\ast)^n \ar[r,"R = \Lambda R' "] & T_N \ar[r] & 1~.
\end{tikzcd}
\end{equation}
We see that $\cG_{\C} = \cG'_{\C}\times \ker \Lambda$, and therefore the two toric varieties are related by $V = V'/\ker \Lambda$. 

\subsubsection*{The dual of $A(X) = G_{\text{q}}$}
Applying this now to our example, we observe that by restricting $m$ to the invariant lattice points, we obtain a polytope $\Delta' \subset M' \subset M$, where the lattice $M'$ is generated by the lattice points $v_0, v_1,v_2$ and the $y_{012}$ and $x_{012}$ marked in~(\ref{eq:Facepicture}).  This is an index $5$ embedding $M' \subset M$ given by
\begin{align}
L & = \begin{pmatrix} v_0\\ v_1 \\ x_{012} \\ y_{013} \end{pmatrix}
= \begin{pmatrix}
-1 & -1 & -1 & -1 \\
~4 & -1 & -1 & -1 \\
~2 & ~0 & -1 & -1 \\
~1 & -1  & ~0 &-1
\end{pmatrix}~.
\end{align}
It is easy to see that the vertices of $\Delta'$ belong to $M'$ since
\begin{align}
\text{Vert}_{\Delta} L^{-1} = \begin{pmatrix}
~1 & ~0 & ~0&   ~0 \\
~0 & ~1& ~0&   ~0 \\
-1 &  -3&~5&   ~0 \\
-2  & -2 &~0& ~5 \\
~2 & ~4  &-5&  -5
\end{pmatrix}
\end{align}
is an integral matrix, and it is also possible to check that a lattice point in $\Delta$ is invariant if and only if it belongs to $M'$.\footnote{To carry our this computation we used the LatticePolytope package in SageMath~\cite{sage:2021ma}.}   Setting $N'$ to be the dual lattice to $M'$, we now obtain the dual polytope $\Delta'^\circ \subset N'_{\R}$, and it turns out to be a lattice polytope with $6$ lattice points and vertices given by the columns of
\begin{align}
\text{Vert}_{\Delta'^\circ} & = 
\begin{pmatrix}
-1 & ~4 & -1  & -1  & -1 \\
-1 & -1  & ~4 & -1  & -1 \\
-1 & ~0 & ~2 & ~0 & -1 \\
-1 & ~1 & ~1 &  -1 & ~0 
\end{pmatrix}~.
\end{align}
Thus, $\Delta' \subset M'$ and $\Delta'^{\circ} \subset N'$ are a reflexive polytope pair, from which we can again construct a pair of GLSMs for the mirror pair of theories.  These turn out to be our example $X$ from section~\ref{ss:holofun}, and its mirror $X^\circ$.  Moreover, we gave the explicit toric morphism associated to the embedding $L : M' \to M$ which shows $\Xb^{\circ} = X^\circ/\ker \Lambda$.

To summarize, $X^\circ$ is a smooth Calabi-Yau hypersurface, and the group $\ker \Lambda \simeq \Z_5$ is a K\"ahler isometry of $X^\circ$.  This action has a non-empty fixed locus, and  quotient is the orbifold space $\Xb^{\circ}$ obtained from a generic mirror quintic $\Xb$ by tuning the complexified K\"ahler parameters to the $G^\circ$--invariant locus.  In the SCFT this is the statement
\begin{align}
\cC[X^\circ]/(G_{\text{q}})^\circ = \cC[\Xb^\circ]~.
\end{align}
This orbifold has additional $80$ marginal (a,c) deformations in the twisted sectors charged with respect to the quantum symmetry of $\cC[X^\circ]/(G_{\text{q}})^\circ$, and by our general orbifold discussion above, this symmetry is $G^\circ$.

\subsubsection*{The mirror of $A(X)$---general story}
Our example is an explicit realization of a  general phenomenon, where mirror symmetry can be used to relate the torsion group $A(X)$ to a global symmetry in the mirror description.   Suppose $X$ and $X^\circ$ are a mirror pair of smooth Calabi-Yau $3$-folds, and $\pi_1(X) = G$ is a finite abelian group.  $X$ has a universal cover $\Xb$ with a group $G$ of K\"ahler isometries such that $\cC[\Xb]$ has $G$ as a global symmetry group on an appropriate locus in the SCFT moduli space.   We denote this restriction by the notation $\cC[\Xb_{G}]$.  If we suppose further that $\Xb$ has a known mirror $\Xb^{\circ}$, we can use the mirror isomorphism to obtain a family of mirror theories $\cC[ (\Xb_{G})^\circ]$ with a global symmetry $G^\circ \simeq G$.\footnote{Our abstruse notation 
$\cC[(\Xb_{G})^\circ]$ denotes the family of SCFTs mirror to the family 
$\cC[\Xb_{G}]$. As the example already shows, it need not correspond to any smooth geometry, and, indeed, it need not have a geometric interpretation at all, since the $G$---invariant locus may not contain a large complex structure point.
}  We then obtain the following diagram relating various SCFTs:
\begin{equation}
\label{eq:Amir}
\begin{tikzcd}[scale=1.2]
\cC[\Xb_G] \ar[rr,->,"\mub"]  \ar[d,"/G"]& &\cS[(\Xb_G)^\circ]  \simeq  \cC[X^\circ]/(G_{\text{q}})^\circ~~~~~ \ar[d, "/G^\circ"]\\
\cC[X] \ar[rr,->,"\mu"]                            &  & \cC[X^\circ]
\arrow[""{ anchor=center, inner sep=0}, "{G_{\text{q}}} = A(X)", from=2-1, to=2-1, out=-45, in=-135, loop, distance=30pt]
\arrow[""{ anchor=center, inner sep=0}, "{(G_{\text{q}})^\circ}", from=2-3, to=2-3, out=-45, in=-135, loop, distance=30pt]
\end{tikzcd}
\end{equation}
Here $\mu$ and $\mub$ denote the mirror maps, and we have the relationships
\begin{align}
G^\circ &= \{ \mub g \mub^{-1}~~~|~~ g\in G\}~, &
(G_{\text{q}})^\circ & = \{ \mu g \mu^{-1}~~~|~~g\in G_{\text{q}}\}~.
\end{align}

In our explicit example we found that the mirror dual of $A(X)$---a non-geometric symmetry of $\cC[X]$---is a geometric symmetry of $\cC[X^\circ]$.   The general structure and the special features of the example lead to a number of additional questions.  For instance, the states charged with respect to $A(X)$ are string winding modes, and intuition from T-duality would suggest that their duals are some ``momentum'' states that could perhaps be represented by eigenmodes of an appropriate Laplace operator defined on $X^\circ$ and graded by representations of the geometric symmetry $(G_{\text{q}})^\circ$.  It would be useful to make this correspondence precise; perhaps it can be explored using recent advances in numeric methods for CY geometry such as~\cite{Ashmore:2021ohf,Ashmore:2023ajy}.   Further work is also needed to study the correspondence when $\pi_1(X)$ is non-abelian, where more sophisticated categorical methods would need to be used to recover the ``upstairs'' SCFTs from $\cC[X]$ and $\cC[X^\circ]$.

\section{Flat gerbes and discrete torsion} \label{s:gerbesdt}
In this section we turn to the situation where $X$ is simply connected but $B(X) = \{H^3(X,\Z)\}_{\text{tor}}$ is non-trivial.
As a working example we can take the CY $X^\circ$ of the previous section with $h^{1,1}(X^\circ) = 21$ and $h^{1,2}(X^\circ)=1$,
for which $B(X^\circ) = \Z_5$.  The computation of $B(X^\circ)$ was carried out in~\cite{Batyrev:2005jc} as part of a study of integral cohomology of CY $3$-fold hypersurfaces in toric varieties.\footnote{In that reference it was observed that hypersurfaces with non-trivial torsion
subgroups are quite sparse, and in the enormous list of $473\,800\,776$ families there are just $32$ examples with torsion subgroups.  Moreover, these come in $16$ mirror pairs, where each pair $(X,X^\circ)$ consists of a manifold with $A(X) \neq 0$ and $B(X) = 0$, while $A(X^\circ) = 0$ and $B(X^\circ) \simeq A(X)$.  The examples of section~\ref{ss:holofun} show that this kind of simple relationship does not hold in general for CY $3$-folds.}  We saw that $B(X^\circ) \neq 0$ means that the theory can be deformed by turning on a non-trivial gerbe, a choice of $\beta_{\text{t}} \in B(X^\circ)$.  The intuition from the path integral suggests that in the large radius limit we might think of a number of distinct complexified K\"ahler cones with continuous parameters $\beta_{\text{f}} + i J \in H^2(X^\circ,\C)$ and distinguished by $\beta_{\text{t}}$~\cite{Aspinwall:1995rb}.  This leads to a number of questions, including:
\begin{enumerate}
\item if we have an orbifold construction as in~(\ref{eq:Amir}), can we understand the choice of $\beta_{\text{t}}$ as a modification of the orbifold $\cC[\Xb^\circ]/G^\circ$?
\item more generally, given a mirror pair $X,X^\circ$, what is a mirror description of $(X^\circ,\beta_{\text{t}} \neq 0)$?
\end{enumerate}
Perhaps the first thought that comes to mind regarding the first question is the introduction of discrete torsion in the orbifold.  This cannot be the case for our example because $H^2(\Z_5,\GU(1)) = 1$.  Another possible answer to the first question can be given in the context of a GLSM construction:  we might guess that there is a choice of discrete $\theta$-angle that could account for the introduction of $\beta_{\text{t}}$.  This possibility was recently discussed in~\cite{Katz:2023zan}, and there are certainly examples that involve GLSM orbifolds where this is indeed realized~\cite{Hori:2011pd}.  In the latter case, these angles arise as follows.  Let $t_\alpha$ be the complexified FI parameters of the GLSM, and suppose that the theory is invariant under an action of  $g\in G$ provided that we transform the $t_\alpha$ by a linear transformation
\begin{align}
g :t_\alpha \to \textstyle\sum_{\beta} R(g)_\alpha^\beta t_\beta~.
\end{align}
By tuning the $t_\alpha$ to the locus preserved by all $g \in G$ we ensure that $G$ is a symmetry of the theory, and we can take the quotient
of the GLSM by $G$.  However, because the $t_\alpha$ are periodic with $t_\alpha \sim t_\alpha + 2\pi N_\alpha$, for $N_\alpha \in \Z$, we can potentially obtain a number of inequivalent solutions by finding integers $T_\alpha(g)$ such that for each $g$ we have
\begin{align}
t_\alpha - \textstyle\sum_{\beta} R(g)_{\alpha}^\beta t_\beta = T_\alpha(g)~,
\end{align}
where $T_\alpha(g)$ is defined modulo the equivalence
\begin{align}
T_\alpha(g) \sim T_\alpha(g) + N_\alpha - \textstyle\sum_{\beta} R(g)_{\alpha}^\beta N_\beta~.
\end{align}
We applied this idea to the specific example of the $\Xb^\circ$ GLSM, where we used the monomial--divisor mirror map to give an explicit action of $G^\circ = \Z_5$ on the complexified FI parameters, and we found that, at least in this example, there are no non-trivial solutions for the $T_\alpha$.  If there is indeed a choice of discrete $\theta$--angle, its appearance must be a more subtle phenomenon.

To tackle the second question we will turn to a specific mirror construction relevant for $X$ and $X^\circ$:  the original Greene-Plesser duality~\cite{Greene:1990ud}, where the SCFTs for $X$ and $X^\circ$ are realized as an isomorphic pair of (2,2) Gepner models.  We will see that $\cC[X^\circ]$ is presented as a $\Z_5\times\Z_5$ orbifold of the quintic theory, and the choice of $\beta_{\text{t}}$ can be identified with the choice of discrete torsion.  Our next steps therefore are to review~\cite{Greene:1990ud} and to generalize those results to include non-trivial discrete torsion phases.

\subsection{Greene-Plesser mirrors via Landau-Ginzburg orbifolds}
Recall that (2,2) Gepner models~\cite{Gepner:1987vz,Gepner:1989gr} are orbifolds of products of (2,2) minimal models taken so as to preserve modular invariance and the $\GUL\times\GUR$ charge integrality in the NS-NS sector necessary for spacetime supersymmetry.\footnote{While we will only discuss the A-series, the extension to all minimal models is known, and there are classification results relevant for string compactification going back to~\cite{Lutken:1988hc}.}

\subsubsection*{A single minimal model}
Rather than working directly with the abstract SCFT, we find it convenient to phrase the construction in terms of a Landau-Ginzburg orbifold (LGO) by using the correspondence between the $A_{d-2}$ minimal model and a (2,2) Landau-Ginzburg theory~\cite{Martinec:1988zu,Vafa:1988uu} with a single chiral superfield $X$ and superpotential interaction $W= X^d$.   This theory is believed to flow to the $\cA_d = \text{A}_{d-2}$ minimal model\footnote{We denote the $\text{A}_{d-2}$ model by $\cA_d$ to avoid the $-2$ offset in the usual minimal model notation.}
with central charges $c = \cb = 3 (1- 2/d)$, and it has a symmetry $G = \Z_d$, which acts non-chirally in the NS-NS sector by $ \ba \cdot X = e^{2\pi i a/d} X$.  The SCFT has a trivial (a,c) ring, while the (c,c) ring is isomorphic to the quotient ring $\cR = \C[X]/\la X^{d-1}\ra$~ determined by the superpotential.  

For any (2,2) SCFT with $c=\cb$ and $d q_{\sleft} ,d q_{\sright} \in \Z$ we define the Poincar{\'e} polynomial as a trace over the Hilbert space of (c,c) operators $\cH^{\text{cc}}$~\cite{Lerche:1989uy}:
\begin{align}
M^{\text{cc}}(t,\tb) = \Tr_{\cH^{\text{cc}}} \left( t^{d J_{0} } \tb^{d \Jb_{0}}\right)~.
\end{align}
For the single minimal model
\begin{align}
M^{\text{cc}}(t,\tb) = 1 + (t\tb) + \cdots + (t\tb)^{d-2} = \frac{ 1 -(t\tb)^{d-1}}{1-(t\tb)}~.
\end{align}
Using the (non-chiral) spectral flow we obtain the Poincar{\'e} polynomial for the RR ground states, which
we denote by $M(t,\tb)$, as
\begin{align}
M(t,\tb) & = (t\tb)^{-dc/6} M^{\text{cc}}(t,\tb)~.
\end{align}

The duality of~\cite{Greene:1990ud} is based on the observation that the minimal model $\cA_{d}$ has a discrete $\Z_d$ symmetry, and taking an orbifold by this symmetry leads to an isomorphic SCFT, denoted by $\cA_d^{\text{orb}}$, with the isomorphism being exactly the reversal of the sign of $\GUL$ charge.  
Using the results of~\cite{Intriligator:1990ua} we obtain the Poincar{\'e} polynomials of the LG orbifold by the symmetry $G$:  
\begin{align}
M_{\text{orb}}(t,\tb) = \frac{1}{d} \sum_{\ba,\bb\in G} M_a^{b}(t,\tb)~,
\end{align}
where the twisted Poincar{\'e} polynomials defined by
\begin{align}
M^{b}_{a} = \Tr_{\cH^{\text{rr}}_{a}} U_a(b) t^{d J_{0}} \tb^{d\Jb_{0}} 
\end{align}
have the explicit expressions\footnote{We recall our notation:  $\zeta_d$ is a primitive $d$-th root of unity, and $[x]$ denotes the  integer part of $x$.}
\begin{align}
M_0^{b} & = (t\tb)^{-dc/6}  \zeta_d^b \frac{ 1-(t\tb\zeta_d^b)^{d-1}}{1-t\tb\zeta_d^b}~,&
M_{a\neq 0}^b & = (t\tb)^{-dc/6} (t\tb^{-1})^{a - d[a/d]} t^{-1}\tb^{d-1}~.
\end{align}
Using these expressions it is easy to see that
\begin{align}
M_{\text{orb}}(t^{-1},\tb) = M(t,\tb)~,
\end{align}
and this is the realization at the level of Poincar{\'e} polynomials of the isomorphism $\cA_d \simeq \cA_d^{\text{orb}}$, which flips the sign of the left-moving R-charge, and identifies the symmetry $\Z_d$ of $\cA_d$ with the quantum symmetry of $\cA_d^{\text{orb}} = \cA_d/\Z_d$.

\subsubsection*{Gepner orbifolds}
Consider now a product of $d$ copies of the LG theory just discussed.  This will have central charge
\begin{align}
c = \cb = 3(d-2)~,
\end{align}
so that for $d=5$ we obtain the $c=\cb = 9$.  Although this is our ultimate interest, we will keep $d$ general with the hope of clarifying aspects of the construction.   With the superpotential
\begin{align}
W = \sum_{i=1}^{d} X_i^d
\end{align}
the theory has a global symmetry group that includes $G_0 = (\Z_d)^{d}$, the group of all phase symmetries with action
\begin{align}
\ba \cdot X_i = \zeta_d^{a_i} X_i~
\end{align}
for $\ba = (a_1,\ldots,a_d)$. $G_0$ contains the subgroup $F_1\simeq \Z_d$ generated by  $\boldsymbol{f}_1 = (1,1,\ldots,1) \in G_0$, and taking the orbifold by $F_1$ leads to a (2,2) theory with integral $\GUL\times\GUR$ charges in the NS-NS sector---for $d=5$ this is the quintic theory at the Gepner point.  From~\cite{Greene:1990ud} we also know that there is a larger subgroup $G \simeq (\Z_d)^{d-1}$ with $F_1 \subset G \subset G_0$, such that the orbifold of the theory by any subgroup of $G$ leads to a (2,2) theory with integral $\GUL\times\GUR$ charges, and for $d=5$ these correspond to different quotients of the quintic and their mirrors.   A way to characterize $G$ is as the largest subgroup of $G_0$ such that all elements $\ba = (a_1,a_2,\ldots,a_d)$ satisfy
\begin{align}
\textstyle\sum_i a_i = 0 \mod d~.
\end{align}
For our purposes it will be useful to characterize $G$ by using an explicit isomorphism between the group $G_0$ and its Pontryagin dual, given by
\begin{align}
\varphi &: G_0 \to G_0^\ast~, &
\varphi & (a_1,\ldots,a_d) = (a_1,\ldots,a_d)~.
\end{align}
For any subgroup $F \subset G_0$ we have the image $\varphi(F) \subset G_0^\ast$.  The group $G$ can then be defined as
\begin{align}
\label{eq:Gdef}
G = \left\{ \ba \in G_0~~|~~ \la \varphi(\boldsymbol{f}_1) , \ba \ra = 0 \right\}~.
\end{align}
More generally, for any $F \subset G_0$ we define $F^\circ \subset G_0$ by
\begin{align}
\label{eq:Fcircdef}
F^\circ = \{ \ba \in G_0 ~~|~~ \la \varphi(\bb), \ba \ra = 0 \quad\text{for all}~ \bb \in F\}~,
\end{align}
and whenever $F$ obeys the inclusions $F_1 \subseteq F \subseteq G$, $F^\circ$ also satisfies the same inclusions.  Thus, the orbifolds by phase symmetries that lead to (2,2) SCFTs with integral $\GUL\times\GUR$ charges come in pairs $F,F^\circ$,  and mirror symmetry is  the isomorphism $\cA_d/F^\circ \simeq \cA_d/F$.

We now review the proof of this isomorphism, using the RR Poincar{\'e} polynomials as a proxy for the full partition function of the SCFT.\footnote{Gepner's results guarantee that the identities established for these generating functions extend to the full partition functions and the SCFT.}  The essential step is to establish the identity
\begin{align}
\label{eq:KeyGP}
(\cA_d^{\text{orb}})^d/\varphi(F)=(\cA_d)^d/F^\circ ~.
\end{align}

Because $G_0^\ast$ is the quantum symmetry of the orbifold theory $(\cA_d^{\text{orb}})^d$, the Poincar{\'e} polynomial for $ (\cA_d^{\text{orb}})^d/\varphi(F)$ is
\begin{align}
Z^{\text{orb}}_{\varphi(F)}(t,\tb) = \frac{1}{|\varphi(F)|} \sum_{\ba^\ast,\bb^\ast\in \varphi(F)} \frac{1}{|G_0|} \sum_{\ba,\bb\in G_0} \zeta_d^{\la \ba^\ast,\ba\ra + \la \bb^\ast,\bb\ra} \prod_{i=1}^d M_{a_i}^{b_i}(t,\tb)~.
\end{align}
Performing the sum on the elements of $\varphi(F)$ and using~(\ref{eq:Fcircdef}), we see that the only terms with non-zero contributions are those with $\ba,\bb\in F^\circ$, and each non-zero term comes with a multiplicity of $|\varphi(F)|^2$.  Since $|\varphi(F)| = |G_0|/|F^\circ|$, we obtain
\begin{align}
Z^{\text{orb}}_{\varphi(F)}(t,\tb) =  \frac{1}{|F^\circ|} \sum_{\ba,\bb\in F^\circ} \prod_{i=1}^d M_{a_i}^{b_i}(t,\tb)~,
\end{align}
but this is exactly the Poincar{\'e} polynomial for $(\cA_d)^d/F^\circ$, so that~(\ref{eq:KeyGP}) holds.  

Mirror symmetry now follows because of the minimal model isomorphism $\cA_d \simeq \cA_d^{\text{orb}}$, and we obtain
\begin{align}
Z_{F}(t^{-1},\tb) = Z_{F^\circ}(t,\tb)~.
\end{align}

\subsection{Quintic quotients} \label{ss:quinticquotients}
Applying this to the case of the quintic with $d=5$, we obtain the following table of Hodge numbers and symmetry actions.
For each symmetry $F \subset G_0$ we list the generators as rows, and the Hodge numbers in the left-hand column are relative to the quotient $(\cA_5)^5/F$.  Of course the quotient with respect to $F^\circ$ yields the reversed Hodge numbers.
\begin{table}[t]
\begin{center}
\begin{tabular}{c|c|c|c}
$h^{1,2},h^{1,1}$  &   $F$     &$F^\circ$ 	&    $K(F) = K(F^\circ)$  \\[3mm]  
$101,1$		    &$F_1 :\begin{pmatrix}1 & 1 & 1 & 1 & 1\end{pmatrix}$   
& $F_1^\circ: \begin{pmatrix} 1 & 1 & 1 & 1 & 1 \\ 4 & 1 & 0 & 0 & 0 \\ 4 & 0 & 1 & 0 &0 \\ 4 & 0 & 0 & 1 & 0\end{pmatrix}$
&  $S_5$
\\[10mm]
$49,5$		    &$F_2 :\begin{pmatrix} 1 & 1 & 1 & 1 & 1 \\ 0 & 0 & 0 & 1 & 4 \end{pmatrix}$   
& $F_2^\circ:\begin{pmatrix} 1 & 1 & 1 & 1 & 1 \\ 4 & 0 & 1 & 0 & 0 \\ 3 & 0 &0 & 1 &1\end{pmatrix}$
& $S_3\times\Z_2$ \\[8mm]
$21,1$		&  $F_3 : \begin{pmatrix} 1 & 1 & 1 & 1 & 1 \\ 0 & 1 & 2 & 3 & 4\end{pmatrix}$ 
& $F_3^\circ:\begin{pmatrix} 1 & 1 & 1 & 1 & 1 \\ 1 & 3 & 1 & 0 & 0 \\ 3 & 1 & 0 & 0 &1\end{pmatrix}$
&
$\Z_5\rtimes\Z_4$ \\[8mm]
$17,21$		& $F_4 :\begin{pmatrix} 1 & 1 & 1 & 1 & 1 \\ 0 & 1 & 1 & 4 & 4 \end{pmatrix}$ 
&$F_4^\circ: \begin{pmatrix} 1 & 1 & 1 & 1 & 1 \\ 0 & 4 & 1 & 0 & 0 \\ 3 & 1 & 0 & 0 &1 \end{pmatrix}$ 
& $\Z_4 \rtimes \Z_2 $ 
\end{tabular}
\label{tab:QGepner}
\caption{Quintic Gepner quotients and their symmetries.  The first column lists the Hodge numbers for the quotient by $F$, and the last column lists the subgroup of the permutation group of the parent theory that survives as a global symmetry in the orbifold.  Notice that $F_1^\circ  = G$---the maximal group of phase symmetries consistent with integral $\GUL\times\GUR$ charges.}
\end{center}
\end{table}

\subsubsection*{A comment on symmetries}
The last column in the table describes a part of the global symmetry of the orbifold theory.  The parent LG theory has a global discrete symmetry $\Gb = S_5 \ltimes G_0 $, with $G_0 =  (\Z_5)^5$.  Taking the quotient by $F$ reduces these symmetries to $\cN(F)/F$, where $\cN(F)$ is the normalizer subgroup
\begin{align}
\cN(F) = \left\{ \gb \in \Gb~~|~~ \gb f \gb^{-1} \in F ~~\text{for all}~~f \in F\right\}~.
\end{align}
Let $R(\sigma) : S_5 \to \GL(5,\Z)$ be the standard $5$-dimensional representation of the symmetric group, so that the group product on $\Gb$ has the form $(\sigma_1, \ba_1) (\sigma_2,\ba_2) = (\sigma_1\sigma_2,\ba_2+\ba_1 R(\sigma_2)) $. Because $S_5$ has trivial center, we find
\begin{align}
\cN(F) = K(F) \ltimes G_0~,
\end{align}
where
\begin{align}
K(F) = \{ \sigma \in S_5~~|~ f R(\sigma) \in F ~~\text{for all} ~ f \in F\}~.
\end{align}
We then see that $\cN(F)/F \simeq K(F) \ltimes G_0/F$~.  We determined $K(F)$ for each of the quotients in the table, and we verified that, as might expected by mirror symmetry, $K(F) \simeq K(F^\circ)$.

What is the  discrete symmetry of the orbifold theory?  A naive guess is that it is simply the product of the symmetry group that survives the quotient, i.e. $\cN(F)/F$, and the quantum symmetry $F^\ast$:
\begin{align}
\Gb_{\text{orb}}(F) \overset{??}{=}  (K(F)  \ltimes G_0/F) \times F^\ast~.
\end{align}
Taking a look at the definitions above, we see that  $F^\ast = G^\ast_0/\varphi(F^\circ)$.  Since we have $K(F) \simeq K(F^\circ)$, the result is almost mirror symmetric, except that the permutation action commutes with the quantum symmetry in our naive guess.  A simple way to obtain a mirror-symmetric expression is to modify the guess to
\begin{align}
\label{eq:globalsym}
\Gb_{\text{orb}}(F) \overset{?}{=} K(F) \ltimes \left( G_0/F \times F^\ast\right) = K(F) \ltimes  \left(G_0/F\times G_0^\ast/\varphi(F^\circ)\right)~.
\end{align}
Let $\ba^\ast,\bb^\ast \in F^\ast$ be a background for the quantum symmetry, so that
\begin{align}
Z_{F}(t,\tb;\ba^\ast,\bb^\ast) = \frac{1}{|F|} \sum_{\ba,\bb\in F} \zeta_5^{\la\ba^\ast,\ba\ra+\la\bb^\ast,\bb\ra} \prod_{i=1}^d M_{a_i}^{b_i}(t,\tb)~.
\end{align}
We then see that $\sigma \in K(F)$ does induce an action on the pair $(\ba^\ast,\bb^\ast)$ as implied by~(\ref{eq:globalsym}).  

As pointed out in~\cite{Tachikawa:2017gyf}, the symmetry of the orbifold theory can be very intricate, and the Gepner quotients, together with their mirrors, provide a rich set of examples where this symmetry structure can be explored in some detail.  A detailed study of some Gepner models has been carried out recently in~\cite{Cordova:2023qei}, with intriguing applications to other SCFTs obtained by deforming away from the Gepner point by marginal deformations.  It should be possible to extend that treatment to other Gepner quotients, such as the ones described here, and to study implications of $\Gb_{\text{orb}}(F)$ and its non-invertible extensions for the deformed theory.  We leave that extension for future work.

\subsubsection*{Discrete torsion}
Returning now to our main thread, we observe that for most of the quotients in our table $H^2(F,\GU(1)) \neq 1$, and there is a possibility of turning on discrete torsion $\varepsilon(\bb,\ba)$, which we parameterize as
\begin{align}
\varepsilon(\bb,\ba)  = \zeta_5^{\lambda(\bb,\ba)}~,
\end{align}
where $\lambda(\bb,\ba)$ is an integer satisfying
\begin{align}
\label{eq:DTproperties}
\lambda(\bb,\ba) & = - \lambda(\ba,\bb) \mod 5~, \nonumber\\
\lambda(\bb_1+\bb_2,\ba) & = \lambda(\bb_1,\ba) + \lambda(\bb_2,\ba) \mod 5~, \nonumber\\
\lambda(\ba,\ba) & = 0 \mod 5~,
\end{align}
The discrete torsion is constrained because the orbifold theory must have integral $\GUL\times\GUR$ charges in the NS-NS sector.  This requires~\cite{Intriligator:1990ua,Kreuzer:1994qp}
\begin{align}
\varepsilon(\boldsymbol{f}_1, \ba) = 1
\end{align}
for all $\ba \in F$, where $\boldsymbol{f}_1$ is the generator of $F_1 \subset F$, or, equivalently,
\begin{align}
\label{eq:DTsusyproperty}
\lambda(\boldsymbol{f}_1,\ba) & = 0 \mod 5\quad\text{for all}\quad \ba \in F~.
\end{align}
  Thus, in fact only the $F_i^{\circ}$ groups in the table allow for a non-trivial discrete torsion, and the possibilities are characterized by
\begin{align}
\Lambda \in \Z_5^3  \subset H^2(F_{1}^\circ,\GU(1))  ~, &&
p \in \Z_5 \subset H^2(F_{2,3,4}^\circ,\GU(1)) ~.
\end{align}
The $125$ possibilities in the first group can be substantially reduced, since many choices turn out to be equivalent up to permutations, and moreover, they naturally come in $\Z_5$ orbits. Still, this leaves us with $9$ non-trivial classes.  On the other hand, for the last three groups we just have a single $\Z_5$ orbit for each, labeled by an integer $p \in \Z/5\Z$.  

Taking the orbifolds with $p \neq 0$, we find that the Hodge pairs $(h^{1,2},h^{1,1})$ change as follows:
\begin{align}
(\cA_5)^5/(F_2^\circ,p\neq 0):&  (5,49)  ~\,\to (23,7)~, \nonumber\\
(\cA_5)^5/(F_3^\circ,p\neq 0):&  (1,21)  ~\,\to (1,21)~, \nonumber\\
(\cA_5)^5/(F_4^\circ,p\neq 0):& (21,17) \to (13,9)~. 
\end{align}
There is exactly one case when the Hodge numbers do not jump, and this is the only situation where the Gepner quotient with $p = 0$ can be deformed to a smooth CY geometry with $B(X)\neq 0$:  in fact, this is the manifold $X^\circ$ with $B(X^\circ) = \Z_5$ that we described above.  For each of the other two cases there is also a smooth CY hypersurface geometry associated to the Gepner orbifold with $p =0$, but neither one has a torsion subgroup in $H^3$.

What are the mirrors for these Gepner orbifolds with $p \neq 0$?  A hint is provided by the Hodge number pairs that arise from the quotients $(\cA_5)^5/(F_1^\circ,\Lambda\neq 0)$. Examining the possibilities in detail, we obtain the Hodge number pairs $(h^{1,2},h^{1,1}) \in \{ (7,23),~(21,1),~(9,13)\}$, which are mirror to those that arise from the $F^\circ_{2,3,4}$ orbifolds with $p \neq 0$.  As we will now show, a generalization of the Greene-Plesser duality allows us to identify the mirror pairs explicitly.

\subsubsection*{Discrete torsion mirrors}
Our goal is to describe the mirror to the quotient $(\cA_5)^5/(F,\lambda)$, with $F = F_{2,3,4}^\circ$, which has the Poincar{\'e} polynomial
\begin{align}
Z_{F,\lambda}(t,\tb) = \frac{1}{|F|} \sum_{\ba,\bb\in F} \zeta_5^{\lambda(\bb,\ba)} \prod_{i=1}^5 M^{b_i}_{a_i}(t,\tb)~,
\end{align}
where $\varepsilon(\bb,\ba) = \zeta_5^{\lambda(\bb,\ba)}$, and $\lambda(\boldsymbol{f}_1,\ba) = 0 \mod 5$ for all $\ba \in F$.

Using the isomorphism $(\cA_d)^d \simeq (\cA_d^{\text{orb}})^d$, we find
\begin{align}
Z_{F,\lambda}(t^{-1},\tb) = Z^{\text{orb}}_{\varphi(F),\lambda}(t,\tb)~,
\end{align}
where 
\begin{align}
Z^{\text{orb}}_{\varphi(F),\lambda}(t,\tb) =
\frac{1}{|\varphi(F)|} \sum_{\ba^\ast,\bb^\ast\in \varphi(F)}  \zeta_5^{\lambda(\bb^\ast,\ba^\ast)} \frac{1}{|G_0|} \sum_{\ba,\bb\in G_0} \zeta_5^{\la \ba^\ast,\ba\ra + \la \bb^\ast,\bb\ra} \prod_{i=1}^5 M_{a_i}^{b_i}(t,\tb)~.
\end{align}
Because there is no discrete torsion between $F_1 \subset F$ and any other element in $F$, we can carry out the sum over $\bb^\ast,\ba^\ast \in F_1$ and arrive at the slightly simpler expression
\begin{align}
Z^{\text{orb}}_{\varphi(F),\lambda}(t,\tb) &=
\frac{|\varphi(F_1)|}{|\varphi(F)|} \sum_{\ba^\ast,\bb^\ast\in \varphi(F)/\varphi(F_1)}  \zeta_5^{\lambda(\bb^\ast,\ba^\ast)} \frac{1}{|G|} \sum_{\ba,\bb\in G} \zeta_5^{\la \ba^\ast,\ba\ra + \la \bb^\ast,\bb\ra} \prod_{i=1}^5 M_{a_i}^{b_i}(t,\tb)~\nonumber\\
& = \frac{1}{|G|} \sum_{\ba,\bb\in G}  S(\bb,\ba) \prod_{i=1}^5 M_{a_i}^{b_i}(t,\tb)~,
\end{align}
where
\begin{align}
S(\bb,\ba)  &= \frac{1}{25}\sum_{\ba^\ast,\bb^\ast\in \varphi(F)/\varphi(F_1)}  \zeta_5^{X(\bb,\ba;\bb^\ast,\ba^\ast)}~, \nonumber\\
X(\bb,\ba;\bb^\ast,\ba^\ast) & = \lambda(\bb^\ast,\ba^\ast) + \la \ba^\ast,\ba\ra + \la \bb^\ast,\bb\ra~.
\end{align}
We will now show that $S(\bb,\ba)$ is a discrete torsion phase for the Gepner orbifold $(\cA_5)^5/G$.  To do this, we specialize to $F = F^\circ_{2,3,4}$, so that $\varphi(F)$ has generators $\boldsymbol{f}_1^\ast, \boldsymbol{f}_2^\ast, \boldsymbol{f}_3^\ast$, with the first one generating the subgroup $\varphi(F_1) \subset \varphi(F)$.  We now expand the elements of $\varphi(F)/\varphi(F_1)$ as
\begin{align}
\ba^\ast &= \ell_2 \boldsymbol{f}_2^\ast + \ell_3 \boldsymbol{f}_3^\ast~,&
\bb^\ast &= m_2 \boldsymbol{f}_2^\ast + m_3 \boldsymbol{f}_3^\ast~,
\end{align}
where $\ell_\alpha,m_\alpha \in \Z / 5\Z$~.  The discrete torsion phase is parameterized by the integer $p$:
\begin{align}
\lambda(\bb^\ast,\ba^\ast) = p(m_2\ell_3-m_3\ell_2) ~.
\end{align}
We also introduce the notation
\begin{align}
\la \ba\ra_{\alpha} = \la \boldsymbol{f}_\alpha^\ast, \ba\ra~,
\end{align}
so that 
\begin{align}
X(\bb,\ba;\bb^\ast,\ba^\ast) & = p(m_2\ell_3-m_3\ell_2)  + \ell_2 \la \ba\ra_2 + \ell_3 \la \ba\ra_3 + m_2 \la \bb \ra_2 +m_3 \la \bb\ra_3~ \nonumber\\
& = m_2 \left( \la \bb\ra_2+ \ell_3 p \right) + m_3\left( \la \bb\ra_3 - \ell_2 p\right) + \ell_2 \la \ba\ra_2 + \ell_3 \la \ba\ra_3~.
\end{align}
Carrying out the sums on $m_2$ and $m_3$, we find
\begin{align}
S(\bb,\ba) = \sum_{\ell_2,\ell_3 =0}^4 \sum_{s_2,s_3 \in 5\Z} \delta(\la \bb\ra_2+ \ell_3 p -s_2) \delta(\la \bb\ra_3 - \ell_2 p-s_3) \zeta_5^{\ell_2 \la \ba\ra_2 + \ell_3 \la \ba\ra_3}~.
\end{align}
When $p = 0$, we carry out the remaining sums on $\ell_\alpha$, and these restrict the elements $\ba,\bb \in F^\circ$, just as in our discussion above.  On the other hand, when $p \neq 0$, let $\pt$ be the multiplicative inverse of $p$ in $\Z/ 5\Z$.  Then the sums on $\ell_2$ and $\ell_3$ only receive contributions from $\ell_2 = \pt \la \bb\ra_3$ and $\ell_3 = -\pt \la \bb\ra_2$, leading to
\begin{align}
S(\bb,\ba) = \zeta_5^{\Lambda_p(\bb,\ba)}~,
\end{align}
with
\begin{align}
\Lambda_p(\bb,\ba) = \pt \left( \la \boldsymbol{f}_3^\ast,\bb\ra \la  \boldsymbol{f}_2^\ast,\ba\ra - \la \boldsymbol{f}_3^\ast,\ba\ra \la  \boldsymbol{f}_2^\ast,\bb\ra\right)~.
\end{align}
This phase satisfies the defining properties of discrete torsion~(\ref{eq:DTproperties}), as well as the integral R-charge constraint~(\ref{eq:DTsusyproperty}) because $\varphi(F) \subset \varphi(G)$, and $\la  \bb^\ast, \boldsymbol{f}_1\ra = 0$ for every $\bb^\ast \in \varphi(G)$.  Thus, we have shown
\begin{align}
Z_{F,p}(t^{-1},\tb) = Z_{G,\Lambda_p}(t,\tb)~,
\end{align}
where $F$ is one of $F^\circ_{2,3,4}$, and $p\neq 0$ specifies the discrete torsion, while $G = F^\circ_1$, and $\Lambda_p$ is the discrete torsion in the mirror description.  Since identical manipulations can be carried out for the full minimal model partition functions, we have established, at least at the level of partition functions, the mirror isomorphism
\begin{align}
(\cA_5)^5/(F,p) \simeq (\cA_5)^5/(G,\Lambda_p)~.
\end{align}
It is straightforward to repeat the argument by starting with the orbifold $(\cA_5)^5/(G,\Lambda)$, where $\Lambda$ is a generic discrete torsion, now labeled by three parameters $p',q',r'\in \Z/5\Z$.  For each such choice, the mirror is again an orbifold of the form $(\cA_5)^5/(F,p)$, where $F$ satisfies the inclusions $F_1 \subset F\subset G$, and both $F$ and $p$ are determined by $p',q',r'$.

\subsection{Discrete torsion and flat gerbes}

By studying the $\Z_5 \times \Z_5$ orbifolds of the quintic Gepner model we observed that turning on discrete torsion for $\Z_5 \times \Z_5$ leads to a jump of Hodge numbers (i.e. a change in the number of (a,c) and/or (c,c) ring elements relative to the orbifold without discrete torsion) for all possibilities with the exception of the quotient by the group $F^\circ_{3}$.  When taken without discrete torsion that orbifold appeared in section 4 of \cite{Batyrev:2005jc} as the Gepner point for the unique CY hypersurface in a toric variety with $B(X) = \Z_5$.  It was conjectured in \cite{Batyrev:2005jc} that the discrete torsion in this case should be in one-to-one correspondence to the flat $B$ field classified by $B(X)$. Our study of the $\Z_5\times \Z_5$ orbifolds provides evidence for this conjecture: the Hodge numbers should not jump if the discrete torsion on the worldsheet construction is identified with flat B-field on the CY3-fold,\footnote{The physical intuition for this is simple:  when $X$ is smooth and at large volume, turning on a flat gerbe amounts to assigning different weights to instanton sums, but it should not alter the spectrum of marginal operators.  The situation is very different when $X$ is singular, as can be seen already in torus orbifolds~\cite{Cheng:2022nso}.} and, indeed, they do not.  For the remaining orbifolds in table~\ref{tab:QGepner} the large radius geometry has a description as a toric hypersurface, but in each case by the classification of~\cite{Batyrev:2005jc} $B(X) = 0$, and thus there is no way to identify the discrete torsion with a flat gerbe; the spectrum jumps as soon as we turn on the discrete torsion.

In this section we will explore the relationship between discrete torsion at the Gepner point and flat gerbes in more detail.  We will argue that while some choices of discrete torsion can be plausibly identified with a flat gerbe over the smooth geometry, in general the set of possible discrete torsions at the Gepner point is larger.  The relationship between discrete torsion and flat gerbes depends on the combinatoric details of the hypersurface in the toric variety.  In some sense this is not surprising, since $B(X)$ is determined by the combinatorics of the toric resolution of the singular ambient space.  It would be nice to understand the relationship more intrinsically in terms of the local geometry of the resolution, but we will leave that study for future work.  Before diving into the further examples, we will discuss the combinatorics of the $\Z_5\times \Z_5$ orbifolds in more detail to motivate a guess for the relationship between discrete torsion and flat gerbes.  We will then check our guess in the examples from the classification of~\cite{Batyrev:2005jc}.

\subsubsection*{Observations on the quintic quotients}
For each of the Gepner quotients $F^\circ_{2,3,4}$ we have the corresponding orbifold of the Fermat hypersurface in $\P^4$, where the action on the projective coordinates of $\P^4$ is encoded by the last two generators of $F^\circ_{2,3,4}$~\cite{Greene:1990ud}.\footnote{The verification that the resolved geometry has precisely the Hodge numbers to match the chiral rings of obtained at the Gepner point was an important step in establishing the mirror duality, and has also been understood and generalized in various studies of stringy Hodge numbers~\cite{Cox:2000vi,MR2359514} and the LG/CY correspondence~\cite{Witten:1993yc}.}  Since the first generator acts trivially by rescaling all of the homogeneous coordinates, we will just denote the ambient toric orbifold variety $V$  by $\P^4 / F^\circ_{2,3,4}$.
We observe that:
\begin{enumerate}[i.]
\item for $\P^4/F^\circ_{3}$ the fixed locus consists of curves in $\P^4$, which intersect the Fermat hypersurface in isolated points;
\item for $\P^4/F^\circ_{2,4}$ the fixed locus includes surfaces, which intersect the hypersurface along curves. 
\end{enumerate}
We have the following facts for a CY $3$-fold hypersurface $X$ in a toric ambient space $V$~\cite{Cox:2000vi,Batyrev:2005jc} with a polytope $\Delta^\circ \subset N_{\R}$.
\begin{enumerate}[1.]
\item There is a toric resolution of singularities  $\widehat{V} \to V$ obtained by refining the fan $\Sigma_V$ through the introduction of  one-dimensional cones $\rho$ for the non-zero lattice points in $\Delta^\circ \subset N_{\R}$, such that the resulting CY variety $\widehat{X} \subset \widehat{V}$ is smooth.
\item The resolution of singular curves in $V$ (and thus of points in $X$) involves $\rho$ contained in the relative interior of codimension $2$ faces $\Theta_2 \subset \Delta^\circ$.
\item The resolution of singular surfaces in $V$ (and thus of curves in $X$) involves $\rho$ contained in the relative interior of codimension $3$ faces $\Theta_3 \subset \Delta^\circ$.
\item When $X \subset V$ is smooth, corollary 3.9 of~\cite{Batyrev:2005jc} determines $B(X)$ as
\begin{align}
B(X) = \Hom\left( \wedge^2 N / \{ N \wedge N''\}, \GU(1)\right)~,
\end{align}
where $N'' \subset N$ is the sublattice generated by the lattice points contained in all faces $\Theta \subset \Delta^\circ$ with codimension $\ge 3$.  It is also shown that $B(X)$ is a cyclic group.
\end{enumerate}
We also have the SCFT perspective on the resolution:  the marginal (a,c) deformations corresponding to the blow-up modes reside in the twisted sector of the orbifold, which are in turn labeled by group elements $\ba \in F$.   If a blow-up mode for the resolution of a singular curve in $X$ resides in such a twisted sector, and the discrete torsion phase $\varepsilon(\bb,\ba)$ is non-trivial, then we expect the spectrum to be affected by turning on discrete torsion.

Taken together, these observations and the quintic results lead to the following guess:  when the discrete torsion phases at the Gepner point  do not affect the (a,c) ring elements associated to $\rho$ contained in $\Theta_3$, we do not expect a jump in the Hodge numbers, and such choices of discrete torsion should correspond to turning on a non-trivial gerbe $\beta_{\text{t}} \in B(X)$ in the smooth geometry.  Next we will analyze several Gepner points for the CY hypersurfaces with $B(X) \neq 0$ that appear in~\cite{Batyrev:2005jc} and examine the fixed loci for the orbifold action on the related weighted projective space.\footnote{Among the $16$ hypersurfaces classified in~\cite{Batyrev:2005jc} there are $5$ cases where the polytope $\Delta$ is a simplex, which guarantees that the mirror theory has $B(X) \neq 0$ and also has a Gepner point.  These are the cases we study in this paper.  The first one is the quotient $\P^4/F^\circ_3$ with $B(X) = \Z_5$ that we considered already, while the remaining $4$ we discuss next.  Of course it may well be that there are other examples in the list with a Gepner point or LGO locus.}    We will see that discrete torsion is more general than the choice of gerbe in $B(X)$.  More precisely, suppose we have a Gepner orbifold with orbifold group $F$, and let $G_{\text{dt}}\subset H^2(F,\GU(1))$ be the group of discrete torsion phases that are consistent with spacetime supersymmetry, i.e. those phases that satisfy~(\ref{eq:DTsusyproperty}).  Our examples suggest that whenever the theory can be deformed to an SCFT $\cC[X,\beta_{\text{t}}]$ based on a smooth geometry $X$, $B(X) \subseteq G_{\text{dt}}$, and in general the inclusion is proper.

\subsubsection*{A Gepner model for $X$ with $B(X) = \Z_3$}
Our first case is the mirror to example 3 from the classification of~\cite{Batyrev:2005jc}.  The Gepner model is realized as a LG orbifold with superpotential 
\begin{equation}
    W = X_1^3 + X_2^3+ X_3^9+ X_4^9+ X_5^9~.
\end{equation}
The projection to integral R-charge requires gauging the R-symmetry
\begin{equation}
     \sigma_{R} \in \Z_9^{R}: (X_1,X_2,X_3,X_4,X_5) \to  (\zeta_9^3X_1,\zeta_9^3X_2,\zeta_9X_3,\zeta_9X_4,\zeta_9X_5)~, 
\end{equation}
and we consider a further quotient by $\Z_3 \times \Z_9$ with generators
\begin{align}
\label{action 1}
    \sigma_2 \in \Z_3 &: (X_1,X_2,X_3,X_4,X_5) \to  (\zeta_9^3X_1,\zeta_9^3X_2,\zeta_9^3X_3,X_4,X_5)~,\\
\label{action 2}
    \sigma_3 \in \Z_9& : (X_1,X_2,X_3,X_4,X_5) \to  (X_1,\zeta_9^6X_2,\zeta_9 X_3,\zeta_9^2X_4,X_5)~. 
\end{align}
The possible discrete torsion is valued in $G_{\text{dt}} =\Z_3$, with the phase given by:
\begin{equation}
\label{eq:DTexample31}
    \varepsilon(\mathbf{b},\mathbf{a})  = \zeta_9^{p\lambda(\mathbf{b},\mathbf{a})},~\qquad \mathbf{b},\mathbf{a} \in \Z_3 \times \Z_9,~\qquad p \in \Z_3~,
\end{equation}
and  $\lambda(\mathbf{b},\mathbf{a})$ is given by:
\begin{equation}
\label{eq:DTexample32}
    \lambda(\bb,\ba) = 3b_2a_3 - 3b_3a_2 \mod 9
\end{equation}
for $\mathbf{b}  = (b_1,b_2,b_3),~\mathbf{a}=(a_1,a_2,a_3) \in F = \Z_9^{R} \times \Z_3 \times \Z_9$~.

The geometric description is based on the orbifold of $V = \P^4_{33111}$ by~(\ref{action 1}) and~(\ref{action 2}), and to study the singularity structure we begin by analyzing the locus of fixed points with respect to the generators of $\Z_3\times\Z_9$.  

For~(\ref{action 1}) this amounts to finding the points in $V$ that satisfy
\begin{equation}
\label{singularity analysis}
    (\zeta_9^3X_1,\zeta_9^3X_2,\zeta_9^3X_3,X_4,X_5) = (\mu^3X_1,\mu^3 X_2,\mu X_3,\mu X_4,\mu X_5)
\end{equation}
for some $\mu \in\C^\ast$.   The fixed locus intersects $(\C^\ast)^4 \subset V$ along the $\C^\ast$ parameterized by $(0,0,0,X_4,X_5)$ with $X_{4,5} \neq 0$, and this is compactified to $\P^1$ in $V$.  Now suppose that $X_4 = X_5 = 0$.  This leaves just two possibilities for the fixed locus:
\begin{enumerate}[1.]
\item
 if $X_4=X_5 = 0$ and $X_3 \neq 0$, then the action on the affine coordinates is
    \begin{equation}
        (\zeta_9^3x_1,\zeta_9^3x_2) = (x_1,x_2)~,
    \end{equation}
which leads to the fixed point $(0,0,1,0,0) \in V$;
\item
 if $X_4 = X_5 = X_3 =0$, then $(X_1,X_2,0,0,0)$ parameterizes another $\P^1\subset V$ fixed by $\sigma_2$.
So, every component of the fixed point locus of $\sigma_2$ has dimension $\le 1$.
\end{enumerate}

Now we turn to (\ref{action 2}).  Consider first the action of the generator $\sigma_3$.  We again see that the fixed locus intersects the dense torus in a $\C^\ast$ orbit, this time parameterized by $(X_1,0,0,0,X_5)$, and this is compactified to a $\P^1_{31} \subset V$ by the inclusion of the $X_1=0$ and $X_5 = 0$ points.   Setting $X_1 = X_5 = 0$, we see that the generic $(\C^\ast)^2$ orbit does not contain any fixed points.  The remaining components of the fixed point locus consist of the orbits  $\P^1_{31}$ parameterized by the points $(0,X_2,0,X_4,0)$, as well as the point $(0,0,X_3,0,0)$.  All in all, we see that all components of the fixed locus of $\sigma_3$  have dimension $\le 1$.  However, there is an additional subtlety, since $\sigma_3^3$ generates a non-trivial $\Z_3 \subset \Z_9$ subgroup which fixes a surface $\P^2_{331}\subset V$, parameterized by $(X_1,X_2,0,0,X_5)$.   By extending this analysis to the remaining elements of $\ba \in \Z_3\times \Z_9$ it is possible to show that the only elements that fix a surface in $V$ have the form $\ba = \sigma_2^k \sigma_3^{3\ell}$, with $\ell = 1,2$.

The divisors introduced in the resolution of singularities correspond to marginal (a,c) operators in the twisted sector of the orbifold CFT.  Applying our general discussion to this example, we see that the twisted sector marginal (a,c) operators associated to resolving curves of singularities in $X$ reside in twisted sectors labeled by $\ba = \sigma_2^k \sigma_3^{3\ell}$, with $\ell = 1,2$.  However, taking a look at the discrete torsion in~(\ref{eq:DTexample31},\ref{eq:DTexample32}) we see that for every such $\ba$, we have $\varepsilon(\bb,\ba) = 1$ for all $\bb \in F$.  Thus,  we do not expect a jump in the Hodge numbers, and performing the computation of the spectrum we find that   $h^{1,2} = 2$ and $h^{1,1} = 38$ for all $p\in G_{\text{dt}}$.

Although in this case the ambient toric orbifold has fixed surfaces, the example is rather like the quintic quotient $\P^4/F^\circ_{3}$:  the choice of discrete torsion appears to be in one-to-one correspondence with $B(X) = G_{\text{dt}} = \Z_3$, and the Hodge numbers do not jump because the twisted sectors that contain the marginal deformations that resolve the fixed surfaces are not affected by the discrete torsion.  The next few examples display more elaborate possibilities, but all of them remain consistent with our proposal laid out above.\footnote{Note that the phase ambiguities in a Landau-Ginzburg orbifold include an additional factor beyond the choice of discrete torsion.  This factor arises as mod 2 ambiguity in the $G$--action on the Ramond-Ramond sector and is related to the Arf invariant and worldsheet spin structures~\cite{Seiberg:1986by,Intriligator:1990ua,Karch:2019lnn}.  When the projection to integral charges has an even order, this additional phase can lead to new possibilities of (2,2) orbifolds.  In this work we set this phase to $+1$ and do not consider it further.  It would be interesting to explore its implications for mirror symmetry and stringy geometry in detail.}

 \subsubsection*{A Gepner model for $B(X) = \Z_2$}
We now turn to example 15 from~\cite{Batyrev:2005jc}.  The Gepner model is realized as LG orbifold with superpotential 
\begin{equation}
    W = X_1^4 + X_2^4+X_3^4+ X_4^8+ X_5^8~,
\end{equation}
and we quotient by $F = \Z_8^R \times \Z_4 \times \Z_4$ with generators
\begin{align}
\sigma_{1} \in \Z_8^{R}:& (X_1,X_2,X_3,X_4,X_5) \to  (\zeta_8^2X_1,\zeta_8^2X_2,\zeta_8^2X_3,\zeta_8X_4,\zeta_8X_5)~,
\nonumber\\   
\sigma_{2} \in \Z_4:& (X_1,X_2,X_3,X_4,X_5) \to  (\zeta_4^2X_1,\zeta_4 X_2,\zeta_4 X_3,X_4,X_5)~, 
\nonumber\\
\sigma_{3} \in \Z_4:& (X_1,X_2,X_3,X_4,X_5) \to  (\zeta_4 X_1,\zeta_4^2 X_2,X_3,\zeta_4 X_4,X_5)~.
\end{align}

The possible discrete torsion, now valued in $G_{\text{dt}} = \Z_4$,  is given by
\begin{equation}
    \varepsilon(\mathbf{b},\mathbf{a})  = \zeta_4^{p\lambda (\mathbf{b},\mathbf{a})}~,
\end{equation}
with
\begin{equation}
 \lambda(\bb,\ba) = b_2a_3  - b_3a_2 \mod 4~.
\end{equation}
The associated geometry is a degree $8$ Fermat hypersurface in $V = \P^4_{11222}/F$, and carrying out the analysis of the fixed locus, we find that there is a $\Z_2 \times \Z_2 \subset F$ subgroup generated by $\sigma_2^2$ and $\sigma_3^2$ that fixes a surface in $V$.  The corresponding twisted sectors are unaffected by taking $p=2$ in the discrete torsion phase, but they are affected when $p=1$ or $p=3$.  Calculating the spectra, we again verify our guess:  the Hodge numbers jump when $p=1,3$, and they do not jump when $p=2$:
\begin{align}
p & = 0~:   h^{1,2} = 5~, h^{1,1} = 29~,\nonumber\\
p & = 1~:   h^{1,2} = 4~, h^{1,1} = 16~,\nonumber\\
p & = 2~:   h^{1,2} = 5~, h^{1,1} = 29~,\nonumber\\
p & = 3~:   h^{1,2} = 4~, h^{1,1} = 16~.
\end{align}
This suggests that $B(X) = \Z_2$ should be identified with the $\Z_2 \subset G_{\text{dt}}$ subgroup  generated by $p=2$.

\subsubsection*{Additional Gepner models with $B(X) = \Z_2$}
There are two more examples from~\cite{Batyrev:2005jc} with an obvious Gepner point.  As their features are much like the case we just analyzed, we will simply summarize the results.
\begin{enumerate}
\item Example 4.  Here the LGO has $F = \Z_8^R\times \Z_4\times \Z_8$ and superpotential
\begin{align}
W = X_1^2 + X_2^8+X_3^8+ X_4^8+ X_5^8~.
\end{align}
The associated geometry is a degree $8$ hypersurface in the ambient toric variety is $V = \P^4_{41111}/F$, and the generators of $F$ are
\begin{align}
\sigma_{1} \in \Z_8^R:& (X_1,X_2,X_3,X_4,X_5) \to  (\zeta_8^4X_1,\zeta_8^2 X_2,\zeta_8^2 X_3,\zeta_8 X_4,\zeta_8X_5)~,\nonumber\\
\sigma_{2} \in \Z_4:& (X_1,X_2,X_3,X_4,X_5) \to  (\zeta_8^4X_1,\zeta_8^2 X_2,\zeta_8^2 X_3,X_4,X_5)~,\nonumber\\
\sigma_{3} \in \Z_8:& (X_1,X_2,X_3,X_4,X_5) \to  (\zeta_8^4 X_1,\zeta_8^3 X_2,X_3,\zeta_8 X_4,X_5)~.
\end{align}
The discrete torsion group is $G_{\text{dt}} = \Z_4$, while the subgroup of $F$ that fixes a surface is $\Z_2 \times \Z_4 \subset F$ generated by $\sigma_2^2$ and $\sigma_3^2$.  Turning on discrete torsion for $\Z_2 \subset \Z_4$ does not affect the twisted sectors that contain the divisors that resolve the fixed surface, and so again, the Hodge numbers only jump when $p = 1,3$:
\begin{align}
p & = 0~:   h^{1,2} = 3~, h^{1,1} = 43~,\nonumber\\
p & = 1~:   h^{1,2} = 3~, h^{1,1} = 19~,\nonumber\\
p & = 2~:   h^{1,2} = 3~, h^{1,1} = 43~,\nonumber\\
p & = 3~:   h^{1,2} = 3~, h^{1,1} = 19~.
\end{align}
This is consistent with the identification $B(X) = \Z_2 \subset G_{\text{dt}}$.

\item Example 7.  The LGO orbifold group is $F = Z_{16}^{R} \times \Z_4 \times\Z_8$, and the superpotential is
\begin{align}
 W = X_1^2 + X_2^4+X_3^8+ X_4^{16}+ X_5^{16}~.
\end{align}
The associated geometry is a degree $16$ hypersurface in $\P^4_{84211}/F$, and $F$ has generators
\begin{align}
\sigma_{1} \in \Z_{16}^R:& (X_1,X_2,X_3,X_4,X_5) \to  (\zeta_{16}^8X_1,\zeta_{16}^4X_2,\zeta_{16}^2X_3,\zeta_{16}X_4,\zeta_{16}X_5)~,\nonumber\\
\sigma_{2} \in \Z_4:& (X_1,X_2,X_3,X_4,X_5) \to  (\zeta_8^4X_1,\zeta_8^2 X_2,\zeta_8^2 X_3,X_4,X_5)~,\nonumber\\
\sigma_{3} \in \Z_8:& (X_1,X_2,X_3,X_4,X_5) \to  ( X_1,\zeta_8^6 X_2,\zeta_8 X_3,\zeta_8 X_4,X_5)~.
\end{align}
The discrete torsion group is $G_{\text{dt}} = \Z_4$, and, as in the previous example, the subgroup of $F$ that fixes a surface is $\Z_2\times\Z_4 \subset \Z_4\times\Z_8$ generated by $\sigma_2^2$ and $\sigma_3^2$.  So, once again we expect that for the discrete torsion subgroup $\Z_2 \subset \Z_4$, which we would like to identify with $B(X)$, the Hodge numbers do note jump. This expectation is borne out by explicit calculation:
\begin{align}
p & = 0~:   h^{1,2} = 3~, h^{1,1} = 75~,\nonumber\\
p & = 1~:   h^{1,2} = 3~, h^{1,1} = 27~,\nonumber\\
p & = 2~:   h^{1,2} = 3~, h^{1,1} = 75~,\nonumber\\
p & = 3~:   h^{1,2} = 3~, h^{1,1} = 27~.
\end{align}
\end{enumerate}
Our examples are consistent with the inclusion $B(X) \subseteq G_{\text{dt}}$, and the jump in the Hodge numbers for $p  \not\in B(X)$ is correlated with resolution of surface singularities in the ambient toric variety.  We also note that for all of these examples we have known mirrors when $p=0$---these are manifolds with $\pi_1(X^\circ) \simeq B(X)$ described in~\cite{Batyrev:2005jc}.  When $p \neq 0$, it is straightforward to extend the construction presented in section~\ref{ss:quinticquotients} to find explicit mirror SCFTs.

\section{Discussion} \label{s:discussion}

The discovery, exploration, and applications of mirror symmetry have enriched a number of fields in physics and mathematics in the last thirty years,\footnote{Pedagogical reviews of some of the most important developments can be found in~\cite{Cox:2000vi,Hori:2003ds,Aspinwall:2004jr,Aspinwall:2009isa}.}
but many fundamental questions remain, and finding the answers to these will lead to significant progress in our understanding of duality in quantum field theory and string theory.   Perhaps the most difficult yet fundamental task that remains is to define precisely in what category of quantum field theories mirror duality operates.\footnote{IVM thanks Ronen Plesser for emphasizing this point in many discussions over the years.  An instructive attempt at tackling this question in the context of linear sigma models was made in~\cite{Aspinwall:2015zia}, but further work is required even in that setting.}   The progress in mirror symmetry at the level of topological field theory---i.e. the homological mirror symmetry program---and the categorical structures that emerge in that approach all suggest that we have much to learn before we can give a fully satisfactory answer to this seemingly basic question.  Our work, based on the earlier efforts in~\cite{Vafa:1994rv,Aspinwall:1994uj,Kreuzer:1994qp,Aspinwall:1995rb,Batyrev:2005jc}, suggests that both orbifolds and flat gerbes should be included in the answer, and we hope that some of the methodology and ideas developed herein will be useful in establishing a suitable framework.

Our study began with~(\ref{eq:torsiongroupsmirror}), which is well-motivated by the intuition that mirror symmetry should exchange even and odd cohomology.  We were able to show that it holds as a consequence of open mirror symmetry and the splitting of Atiyah-Hirzebruch sequences for the torsion K-groups, but with the assumption that every element in $A(X)$ is of odd order.
Since we also have examples where $A(X)$ has even-order elements, but the sequences nevertheless split, it is clear that the result holds beyond what we have been able to prove.  Is it true that for any CY threefold 
\begin{align}
\{K^0(X)\}_{\text{tor}} = A(X) \oplus B(X)^\ast~?
\end{align}

Another natural follow-up to our study would be to consider the relationship between discrete torsion at Gepner points (and more generally Landau-Ginzburg orbifold loci) and $B(X)$ more closely and systematically.  Some efforts in this direction have been made many years ago.  For example, there is a classification of all Landau-Ginzburg orbifolds with discrete torsion~\cite{Kreuzer:1994qp}, but the results obtained therein are limited to the computation of dimensions of (a,c) and (c,c) rings.  Those findings are already intriguing: for instance, there are many entries with relatively small dimensions, and even one example without marginal (2,2) deformations!  It is quite likely that some of these are related to Calabi-Yau manifolds with small Hodge numbers described in~\cite{Candelas:2016fdy}, and the relation probably involves identification of $B(X)$ with a subgroup of the discrete torsion.   It should be useful to study this class of examples along the lines developed in this work by focusing on the relationship between the solvable CFT locus and the resolved geometry.  Such an effort will lead to a better understanding of aspects of the problem that appear obscure (at least to us):  for instance, how universal is the relationship between resolution of fixed curves in the Calabi-Yau geometry and the jumping in the Hodge numbers in the presence of discrete torsion?  First steps in this direction involve re-examining the classification of~\cite{Kreuzer:1994qp} with an eye to geometric interpretation, and also a closer study of the examples in~\cite{Candelas:2016fdy,Constantin:2016xlj} to determine the $A(X)$ and $B(X)$ torsion groups.  It will also be useful to study self-mirror examples, such $X_2$ and $X_3$ described in section~\ref{ss:holofun}.

We also envision a more detailed study of mirror symmetry in the presence of discrete torsion at special orbifold points and related flat gerbes over a smooth geometry.  It will be useful to construct additional examples of mirror pairs with flat gerbes and to elucidate the relationship between the geometries on both sides of the mirror.  In the context of Gepner models we have shown how to construct mirror pairs with the inclusion of discrete torsion.  Can this be extended to other mirror constructions, such as that of Berglund-H\"ubsch for Landau-Ginzburg orbifolds~\cite{Berglund:1991pp}?

There are also more general and perhaps more speculative questions concerning the SCFT associated to  a smooth CY $X$ with non-zero $A(X)$ or $B(X)$:  
\begin{enumerate}
\item Can we identify a flat gerbe on $X$ with a choice of discrete torsion for an orbifold at some point in the moduli space?  
\item If $X$ has multiple orbifold points in the moduli space, what are the relations between the discrete torsions at various points?  
\item Is there always a ``parent'' geometry $\Xb$ with $A(\Xb) = B(\Xb) = 0$, such that on a suitable locus in the moduli space of $\cC[\Xb]$ we can obtain $\cC[X,\beta_{\text{t}}]$ by an orbifold of $\cC[\Xb]$?
\item We have seen that given a mirror pair $X$, $X^\circ$,  there is in general a finite set of SCFTs $\cC[Y^\circ,\beta^\circ_{\text{t}}]$ that are mirror to the discrete family $\cC[X,\beta_{\text{t}}]$, $\beta_{\text{t}} \in B(X)$, with $\cC[X,0] \simeq \cC[X^\circ] = \cC[Y^\circ,0]$.  Which $\cC[Y^\circ,\beta^\circ_{\text{t}}]$ have a geometric description?  Can we describe the discrete family $\cC[Y^\circ,\beta^\circ_{\text{t}}]$ directly, without appealing to the mirror?
\item There are additional topological invariants encoded in the cup product on $\{H(X,\Z)\}_{\text{tor}}$.  Do these have an interpretation in the SCFT $\cC[X]$ and its mirror?

\item Can we give an intrinsic SCFT definition of the groups $A(X)$ and $B(X)$ without reference to a geometry?\footnote{We thank Paul Aspinwall for this question.}
\end{enumerate}

Coming back to our discussion of $A(X)$ and its relation to a quantum symmetry of the SCFT, we can also imagine extending the analysis to the situation where $\pi_1(X)$ is non-abelian.  It will be interesting to understand the categorical symmetry structure in such a situation and to relate it to the geometry and topology of $X$, as well as to interpret it on the mirror $X^\circ$.  This is one of the directions where these old questions can benefit from the relatively new categorical symmetry methods, and at the same time provide concrete examples to develop further intuition for these remarkable structures.  Our work also motivates a closer study of the relationship between discrete torsion and gauged linear sigma models:  are there useful ways to lift this phase to the UV description?  A related point concerns the suggestions---see, e.g.~\cite{Aspinwall:1995rb,Hori:2011pd,Katz:2023zan}, that certain flat gerbes can be interpreted as a discrete $\theta$-angle in the gauged linear sigma model.  It would be useful to study these notions systematically, as they will certainly have important ramifications for explicit mirror constructions.

We concentrated almost exclusively on the worldsheet SCFT, but the structures we discussed have important implications for the effective spacetime theory.  At the most basic level, we expect that $A(X)$ should be interpreted as a spacetime gauge symmetry since by our arguments $A(X)$ is identified with a global symmetry in the worldsheet theory.  On the contrary, $B(X)$ does not have such an interpretation---this is not in conflict with mirror symmetry because, as we argued, mirror symmetry does not exchange $A(X)$ and $B(X)$ in any natural fashion.  Instead, $B(X)$ labels the non-trivial gerbe backgrounds that can be supported on $X$, and turning on a non-trivial gerbe $\beta_{\text{t}} \subset B(X)$ should affect the geometry of $\cM_{\text{ac}}(X)$---the moduli space of vector multiplets in a IIA compactification on $X$.  What, precisely, are the consequences of these features for the effective theory?  There should also be consequences for the spectrum of BPS and non-BPS D-brane states, and the study of the interplay between twisted K-theory and the Freed-Witten anomaly in these concrete examples will surely have important general lessons and perhaps shed light on open problems concerning the spectrum of branes in the presence of topologically non-trivial fluxes. 

The most speculative further directions concern generalizations to higher dimensions.  What are the consequences of these discrete topological features for $\text{G}_2$ mirror symmetry or for mirror symmetry of CY $4$-folds?   While the former has a reasonable spacetime interpretation as a $3$-dimensional string compactification, the worldsheet physics is more difficult, since it involves a theory without extended supersymmetry.  For the latter, although the worldsheet theory is still a (2,2) SCFT, the spacetime interpretation is subtle: it is a ``compactification'' to 1+1 dimensions with a non-trivial tadpole for the Ramond-Ramond flux.

\bibliographystyle{utphys}
\bibliography{newref}

\end{document}